\documentclass[twocolumn,trackchanges]{aastex631} 

\usepackage{graphicx}
\usepackage{subfigure}
\usepackage{paralist}
\usepackage{amssymb}
\usepackage{bm}
\usepackage{bbding}
\usepackage[fleqn]{amsmath}
\usepackage{multirow}
\hypersetup{linkcolor=red,citecolor=blue,filecolor=cyan,urlcolor=magenta}

\shorttitle{Orbit flips}
\shortauthors{Lei}

\graphicspath{{./}{figures/}}

\begin{document}

\title{A systematic study about orbit flips of test particles caused by eccentric von Zeipel--Lidov--Kozai effects}

\correspondingauthor{Hanlun Lei}
\email{leihl@nju.edu.cn}

\author{Hanlun Lei}
\affiliation{School of Astronomy and Space Science, Nanjing University, Nanjing 210023, China}
\affiliation{Key Laboratory of Modern Astronomy and Astrophysics in Ministry of Education, Nanjing University, Nanjing 210023, China}



\begin{abstract}
The problem of orbit flips caused by eccentric von Zeipel--Lidov--Kozai effects is systematically investigated by means of three approaches, including Poincar\'e sections, dynamical system theory (periodic orbits and invariant manifolds) and perturbation treatments. Poincar\'e sections show that orbit flips are due to the existence of islands of libration centered at inclination of $90^{\circ}$, dynamical system theory shows that orbit flips are due to the existence of polar periodic orbits and invariant manifolds, and perturbative treatments indicate that orbit flips are due to the libration of a certain critical argument. Using these approaches, the boundaries of flipping regions in the entire parameter space are produced and they are in excellent agreement with each other. Through analysis, the essence of flipping orbits is reached: (a) flipping orbits are a kind of quasi-periodic trajectories around polar periodic orbits and invariant manifolds at the same level of Hamiltonian provide boundaries of flipping regions, and (b) flipping orbits are a kind of resonant trajectories and resonant width measures the size of flipping regions.
\end{abstract}



\section{Introduction}
\label{Sect1}

Hierarchical three-body configurations are common in various scales of astrophysical systems \citep{naoz2016eccentric}. Under the test-particle approximation, \citet{lidov1962evolution} studied the orbital evolution of artificial satellites under the gravitational perturbations coming from the Sun and Moon, and \citet{kozai1962secular} investigated the secular effects of Jupiter's gravitational perturbations on inclined asteroids in the Solar System. In these two works, the orbits of the perturber are assumed to be circular and thus the secular potential is axisymmetric. The longitude of ascending node $\Omega$ becomes a cycle variable in the associated Hamiltonian model, thus the vertical angular momentum conserves in the long-term evolution, leading to coupled evolutions between eccentricity and inclination for the test particle's orbit. Due to the conservation of vertical angular momentum, the orbits cannot flip under the test-particle and circular approximation. Such an effect of exciting particle's eccentricity and/or inclination under the circular restricted hierarchical three-body problem is referred to as the standard Kozai--Lidov mechanism \citep{lithwick2011eccentric}. Recently, \citet{ito2019lidov} pointed out that von Zeipel (1910) has performed a pioneering work using a similar analysis of this subject and thus they suggested to refer to such a mechanism as the ``von Zeipel--Lidov--Kozai" effect.

If the circular approximation of perturber's orbit is relaxed (i.e., the perturber is assumed to move on a fixed elliptic orbit), the eccentric effect appears under the octupole level of approximation. Under this model, the particle's vertical angular momentum $H=G\cos{i}$ is no longer a constant, thus its Kozai oscillations are modulated on longer timescales \citep{li2014eccentricity}. Particularly, the particle's orbit can flip between prograde and retrograde, and in the process the eccentricity can reach a value close to unity. \citet{lithwick2011eccentric} referred to this effect as the eccentric Kozai mechanism, which can be applied to a vast variety of astrophysical systems \citep{naoz2016eccentric, shevchenko2016lidov}. Following the suggestion given by \citet{ito2019lidov}, we refer to such a mechanism as the eccentric von Zeipel--Lidov--Kozai effect in the current work. \citet{antognini2015timescales} found that the timescale of von Zeipel--Lidov--Kozai oscillations at the quadrupole order is
\begin{equation}
t_{\rm quad} \approx \frac{16}{15}\frac{a_p^3}{a^{3/2}}\sqrt{\frac{m_0}{{\cal G}m_p^2}}\left(1-e_p^2\right)^{3/2}
\end{equation}
where ${\cal G}$ is the universal gravitational constant, $a$ and $a_p$ are the semimajor axes of the test particle and perturber, $e$ and $e_p$ are the eccentricities of the test particle and the perturber, $m_0$ and $m_p$ are the mass of the central star and the perturber. In addition, \citet{antognini2015timescales} showed that the timescale of eccentric von Zeipel--Lidov--Kozai oscillations at the octupole order is proportional to ${\epsilon}^{-1/2}$ where ${\epsilon}$ measures the significance of octupole-order perturbation relative to the quadrupole-order term (see the detailed expression of ${\epsilon}$ in the main body of this work).

About the topic of orbit flips caused by eccentric von Zeipel--Lidov--Kozai effects, \citet{katz2011long} found a new constant of motion by averaging the double-averaged secular equations of motion over the Kozai--Lidov cycles and they derived an analytical criterion for orbit flipping. In the test-particle limit at the octupole level of approximation, \citet{li2014} studied the chaotic and quasi-periodic evolutions under the eccentric von Zeipel--Lidov--Kozai effect (called Kozai--Lidov mechanism in their work) by analysing surfaces of section and Lyapunov exponents. When the mutual inclination between the inner and the outer binary is high ($40^{\circ}-140^{\circ}$), the inner orbit exhibits striking features due to the long-timescale modulations of Kozai--Lidiv cycles \citep{katz2011long}. The striking features include, for example, flipping between prograde and retrograde, significant excitation of eccentricity and/or inclination, and even chaotic behaviors. \citet{li2014eccentricity} denoted the orbit flips happening with inclinations in $[40^{\circ},140^{\circ}]$ as the first type, corresponding to the low-eccentricity high-inclination case (called LeHi for short), and they demonstrated a second type of orbit flipping which starts from an almost coplanar configuration when the eccentricity of the test particle is high, corresponding to the high-eccentricity low-inclination case (called HeLi for short). As for the HeLi case of orbit flips, the flip criterion and the flip timescale are dependent on the initial orbital parameters \citep{li2014eccentricity}. Regarding the HeLi case, \citet{petrovich2015hot} studied the possibility that hot Jupiters are formed from coplanar high-eccentricity migration (called CHEM for short). Recently, \citet{sidorenko2018eccentric} demonstrated that the orbit flips of the LeHi case caused by eccentric von Zeipel--Lidov--Kozai effects can be interpreted as a resonance phenomenon (it is noted that the HeLi case is not covered in their interpretation). Additionally, some non-gravitational factors and high-order truncations of third-body perturbation could influence the conventional von Zeipel--Lidov--Kozai effects. In this context, \citet{will2017orbital} studied the problem of orbit flips under the relativistic effects and third-body effects to the hexadecapole order, and they concluded that for most cases the relativistic effects and the hexadecapole perturbations have negligible influences upon the orbital flips found at the octupole order but, for equal-mass binaries (in this case the octupole terms vanish), hexadecapole perturbations can alone cause orbital flips.

Furthermore, if the test-particle approximation is relaxed (i.e., the test particle is replaced by a massive object), it becomes the non-restricted hierarchical version. Under such a `general' configuration, the octupole-order dynamical model can be seen in \citet{naoz2011hot} and \citet{naoz2013secular}. Caused by eccentric von Zeipel--Lidov--Kozai effects, the planets' orbits can flip between prograde and retrograde with respect to the invariant plane of system. Naturally, such a mechanism can be used to explain the formation of hot Jupiters on retrograde orbits by combining with tidal friction \citep{naoz2011hot, naoz2012formation}. In a recent review, \citet{naoz2016eccentric} provided varieties of applications of the eccentric von Zeipel--Lidov--Kozai effect to solar system dynamics, planetary systems, stellar systems, exoplanetary systems and galaxies etc.

About the problem of orbit flips caused by eccentric von Zeipel--Lidov--Kozai effects, the following questions are still not very clear. Initially, what is the boundaries of flipping orbits in the entire $(e_0,i_0)$ space? Secondly, what's the particular mechanism that dominates the boundaries and what is the underlying mechanism that governs the phenomenon of orbit flipping? Finally, an ultimate motivation for us is to know what the essence of flipping orbits is. To clear these doubts, we numerically explore the flipping regions in the $(e_0,i_0)$ space. Then, three approaches including Poincar\'e surfaces of section, dynamical system theory and perturbative treatments are adopted to explore the numerical structures of flipping regions and, in particular, boundaries of flipping regions are produced by different approaches. From the viewpoint of Poincar\'e sections, orbit flips are due to the existence of islands of libration centered at $i=90^{\circ}$. From the viewpoint of dynamical system theory, orbit flips are due to the existence of polar periodic orbits and invariant manifolds. From the aspect of perturbation treatments, orbit flips are due to the libration of a certain critical argument \citep{lithwick2011eccentric, sidorenko2018eccentric}. The results (boundaries of flipping regions) obtained by these three approaches agree well with each other and, particularly, we are guided to the essence of flipping orbits caused by eccentric von Zeipel--Lidov--Kozai effects: flipping orbits are a kind of quasi-periodic (or resonant) trajectories around the polar periodic orbits.

The remaining part of this work is organized as follows. In Section \ref{Sect2}, the basic dynamical model is introduced under the test-particle octupole approximation. Numerical explorations about flipping regions in the eccentricity--inclination space are performed in Section \ref{Sect3}. The numerical structures of flipping regions are systematically explored through three approaches: Poincar\'e surfaces of section in Section \ref{Sect4}, dynamical system theory (periodic orbits and invariant manifolds) in Section \ref{Sect5} and perturbative treatments in Section \ref{Sect6}. Finally, conclusions and discussions of this work are given in Section \ref{Sect7}.

\section{Hamiltonian model}
\label{Sect2}

In this work, we consider such a hierarchical three-body configuration in which an inner test particle is orbiting around a central star in the presence of a distant massive planet. The configuration is in accordance with the ones considered in previous works \citep{li2014, li2014eccentricity, katz2011long, lithwick2011eccentric, sidorenko2018eccentric}. In the test-particle limit, the perturber is moving on a fixed Keplerian orbit and the test particle is moving on a perturbed Keplerian orbit around the central star. Their orbits are described by the semimajor axis $a(a_p)$, eccentricity $e(e_p)$, inclination $i(i_p)$, longitude of ascending node $\Omega(\Omega_p)$, argument of pericenter $\omega(\omega_p)$ and mean anomaly $M(M_p)$. The longitude of pericenter and mean anomaly are defined by $\varpi = \Omega + \omega$ and $\lambda = M + \varpi$ for the test particle and $\varpi_p = \Omega_p + \omega_p$ and $\lambda_p = M_p + \varpi_p$ for the perturber. Unless otherwise stated, in the entire work we take the variables with subscript $p$ for the perturber and the ones without subscript $p$ for the test particle.

For convenience, we choose an inertial right-handed coordinate system originated at the central star, with the orbit of the perturber (i.e., the invariant plane of the considered system) as the fundamental plane, the $x$-axis and $z$-axis along the eccentricity vector and angular momentum vector of the perturber, respectively. Under such a reference frame, both the inclination of the perturber and longitude of pericenter are zero, i.e., $i_p =0$ and $\varpi_p = 0$.

In hierarchical configurations, the semimajor axis ratio between the test particle and the perturber ($\alpha = a/a_p$) is a small parameter and the Hamiltonian of the system can be expanded as a power series of $\alpha$ \citep{harrington1968dynamical, harrington1969stellar}. In long-term evolutions, the terms containing the short-period angles $M$ and $M_p$ can be removed from the Hamiltonian by means of phase averaging \citep{ford2000secular, naoz2013secular}. Please refer to \citet{lei2021} for the explicit expression of the double-averaged Hamiltonian truncated at an arbitrary order in $\alpha$. The phase-averaging process is known as the secular approximation \citep{naoz2016eccentric}, which corresponds to the lowest-order perturbation treatments \citep{hori1966theory, deprit1969canonical}. Under the averaged model, the semimajor axis of the test particle is conserved and its eccentricity and inclination are coupled in evolution on timescales much longer than the orbital period when the inclination is greater than $39.2^{\circ}$ \citep{lidov1962evolution, kozai1962secular, shevchenko2016lidov}.

For convenience, let us introduce the (normalized) Delaunay variables as follows \citep{morbidelli2002modern, lithwick2011eccentric}:
\begin{equation}\label{Eq1}
\begin{aligned}
g &= \omega ,\quad G = \sqrt {1 - {e^2}},\\
h &= \Omega ,\quad H = G\cos i.
\end{aligned}
\end{equation}
In terms of Delaunay variables, the (double-averaged and scaled) Hamiltonian, truncated at the octupole-level approximation, can be written as \citep{lithwick2011eccentric, li2014eccentricity, li2014, naoz2016eccentric}
\begin{equation}\label{Eq2}
{\cal H}\left( g, G, h, H \right) =  - F_{\rm quad}\left( g, G, H \right) - \epsilon F_{\rm oct}\left( g, G, h, H \right),
\end{equation}
where the quadrupole-level term $F_{\rm quad}\left( g, G, H \right)$ is
\begin{equation}
\begin{aligned}
{F_{\rm quad}} =&  - \frac{1}{2}\left( {1 - {G^2}} \right) + \frac{{{H^2}}}{{{G^2}}} + \frac{3}{2}\left( {\frac{1}{{{G^2}}} - 1} \right){H^2}\\
&+ \frac{5}{2}\left( {1 - {G^2} + {H^2} - \frac{{{H^2}}}{{{G^2}}}} \right)\cos \left( {2g} \right),
\end{aligned}
\end{equation}
the octupole-level term $F_{\rm oct}\left( g, G, h, H \right)$ is
\begin{small}
\begin{equation}
\begin{aligned}
{F_{\rm oct}} &= \frac{5}{{16}}\left[ {\sqrt {1 - {G^2}}  + \frac{3}{4}{{\left( {1 - {G^2}} \right)}^{3/2}}} \right] \\
& \times \left[ {\left( {1 - 11\frac{H}{G} - 5\frac{{{H^2}}}{{{G^2}}} + 15\frac{{{H^3}}}{{{G^3}}}} \right)\cos \left( {g - h} \right)} \right.\\
&\left. { + \left( {1 + 11\frac{H}{G} - 5\frac{{{H^2}}}{{{G^2}}} - 15\frac{{{H^3}}}{{{G^3}}}} \right)\cos \left( {g + h} \right)} \right]\\
& - \frac{{175}}{{64}}{\left( {1 - {G^2}} \right)^{3/2}}\left[ {\left( {1 - \frac{H}{G} - \frac{{{H^2}}}{{{G^2}}} + \frac{{{H^3}}}{{{G^3}}}} \right)\cos \left( {3g - h} \right)} \right.\\
&\left. { + \left( {1 + \frac{H}{G} - \frac{{{H^2}}}{{{G^2}}} - \frac{{{H^3}}}{{{G^3}}}} \right)\cos \left( {3g + h} \right)} \right]
\end{aligned}
\end{equation}
\end{small}
and the factor $\epsilon$, standing for the significance of the octupole-level term, is given by
\begin{equation}
\epsilon  = \frac{a}{{{a_p}}}\frac{{{e_p}}}{{1 - e_p^2}}.
\end{equation}
Evidently, the contribution of the octupole-order term increases with the semimajor axis ratio $\alpha$ and the eccentricity of the perturber's orbit $e_p$. Second-order corrections to the double-averaged Hamiltonian can be referred to \citet{luo2016double}, \citet{lei2018modified} and \citet{lei2019semi}.

The Hamiltonian given by equation (\ref{Eq2}) determines a two-degree-of-freedom dynamical model with $g$ and $h$ as the angular coordinates and it holds the following symmetric properties \citep{sidorenko2018eccentric}:
\begin{equation}
\begin{aligned}
{\cal H} (G, H, g, h) &= {\cal H} (G, -H, 2\pi - g, h) \\
&= {\cal H} (G, -H, g, 2\pi - h) \\
&= {\cal H} (G,H,\pi + g,\pi + h)
\end{aligned}
\end{equation}
According to the Hamiltonian canonical relations, the equations of motion can be expressed by \citep{morbidelli2002modern}
\begin{equation}\label{Eq3}
\begin{aligned}
\frac{{{\rm d} g}}{{{\rm d} t}} = \frac{{\partial {\cal H}}}{{\partial G}},\quad \frac{{{\rm d} G}}{{{\rm d} t}} =  - \frac{{\partial {\cal H}}}{{\partial g}},\\
\frac{{{\rm d} h}}{{{\rm d} t}} = \frac{{\partial {\cal H}}}{{\partial H}},\quad \frac{{{\rm d} H}}{{{\rm d} t}} =  - \frac{{\partial {\cal H}}}{{\partial h}}.
\end{aligned}
\end{equation}
An alternative description for the equations of motion under the octupole-order approximation can be written in the vectorial form, see for instance \citet{katz2011long} and \citet{luo2016double} for more details.

\section{Numerical explorations of orbit flips}
\label{Sect3}

The equations of motion presented in the previous section are numerically integrated in order to explore the regions of orbit flips in the (initial) parameter space $(e_0, i_0)$ with given initial angles $(\Omega_0,\omega_0)$. In practical simulations, we assume the starting points located at the apsidal line with $\varpi_0 = 0$ or $\varpi_0 = \pi$. To study the eccentric von Zeipel--Lidov--Kozai effects under the octupole-level approximation, we only deal with the rotating Kozai cycles. In a certain rotating Kozai cycle, the argument of pericenter $\omega$ can change from $0$ to $2\pi$. In practice, we restrict us to the initial argument of pericenter at zero \footnote{This choice of $\omega_0$ is also adopted by \citet{lithwick2011eccentric} and \citet{li2014eccentricity} in their numerical simulations. However, different choice of $\omega_0$ may produce different results.}, and the initial longitude of ascending node is taken as $\Omega_0 = \pi$ (case I) and $\Omega_0 = 0$ (case II).  It is noted that the initial angles $\omega_0 = 0$ and $\Omega_0 = \pi$ (corresponding to case I) are used by \citet{li2014eccentricity} in their numerical studies. Additionally, numerical explorations about the boundaries of orbit flips are performed by \citet{lithwick2011eccentric} in the eccentricity space $e_0 \in [0,0.5]$.

\begin{figure*}
\centering
\includegraphics[width=0.45\textwidth]{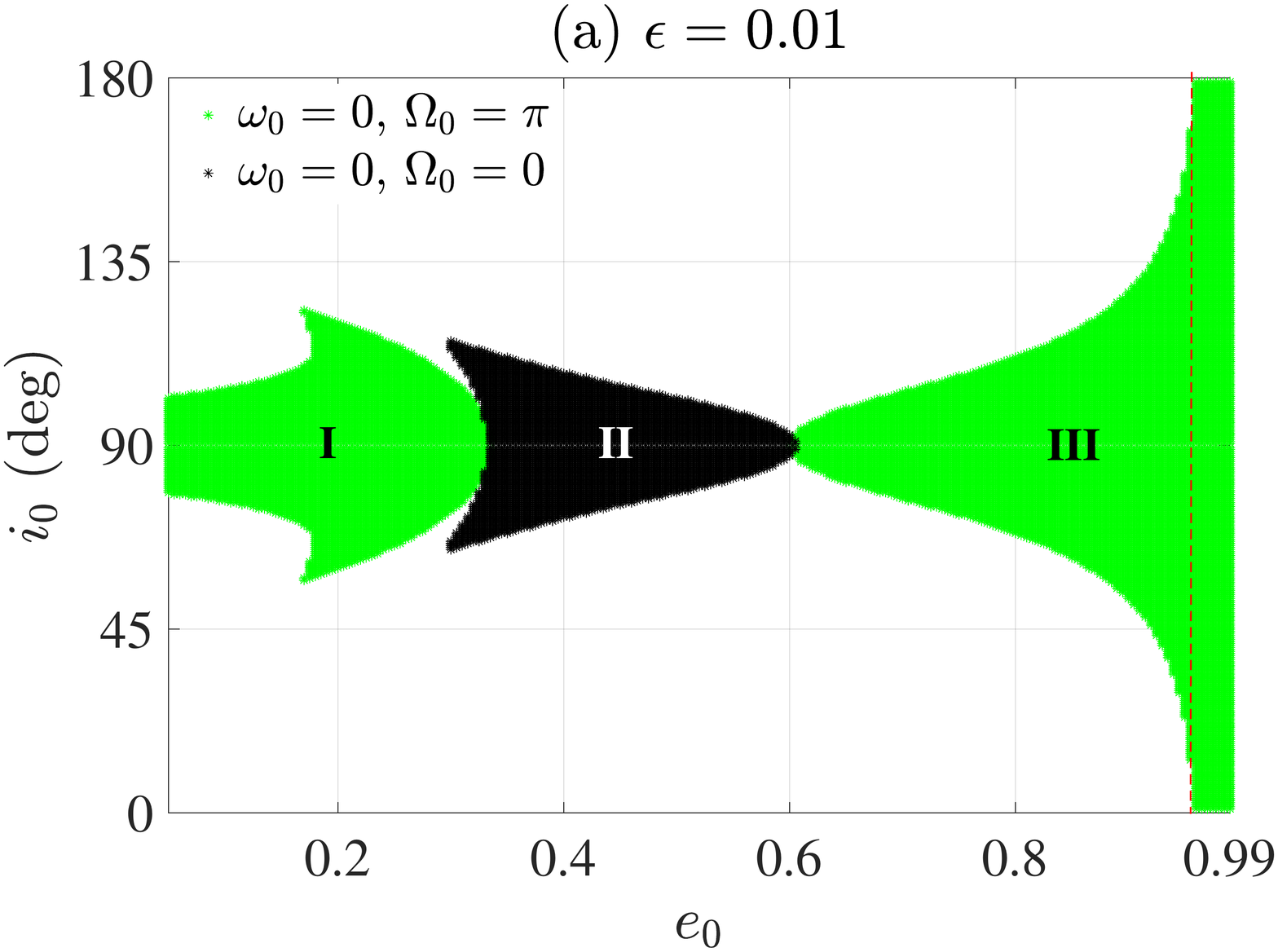}
\includegraphics[width=0.45\textwidth]{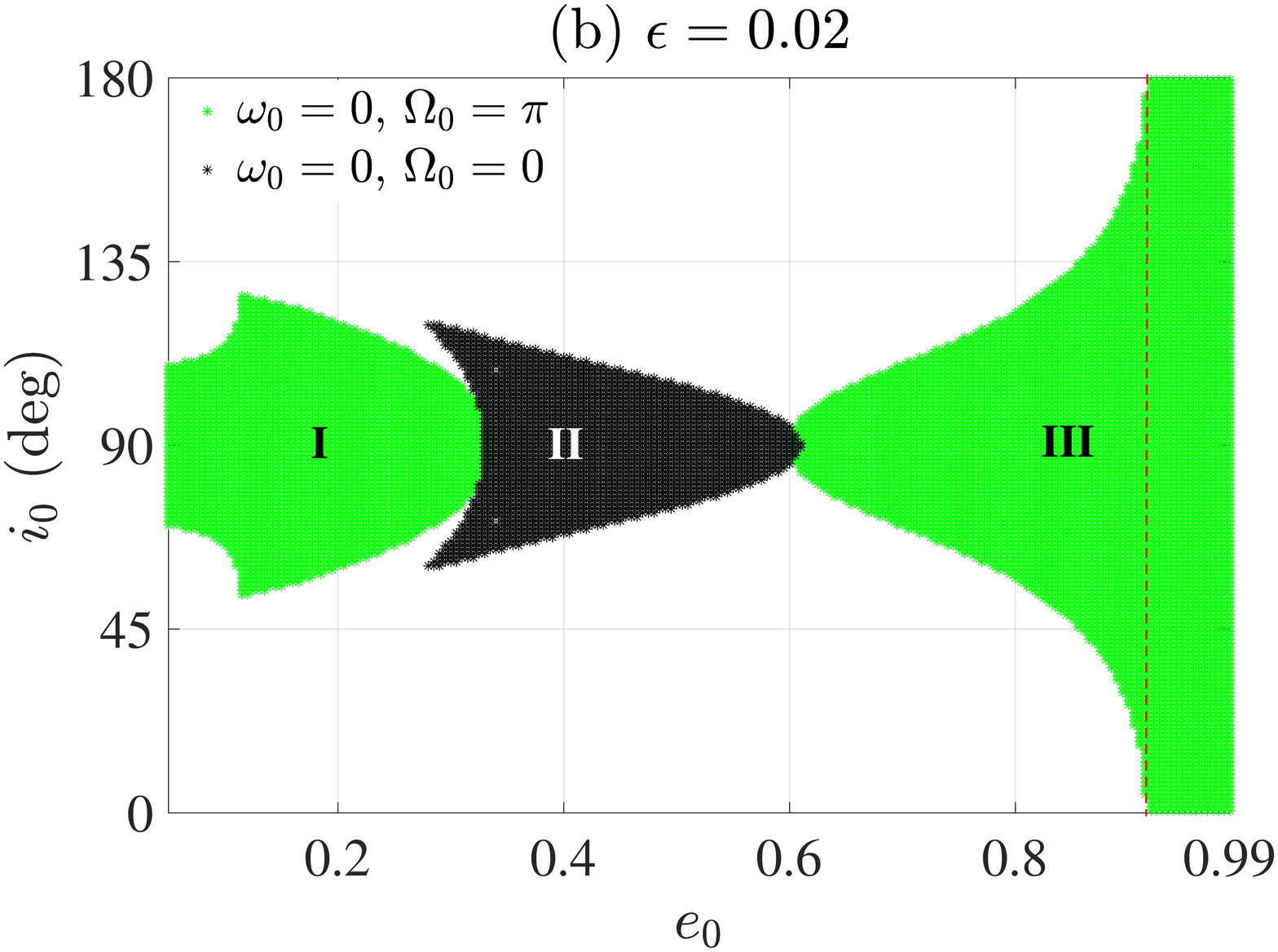}
\caption{Flipping regions in the entire $(e_0,i_0)$ space. The dynamical models are characterized by $\epsilon = 0.01$ (\emph{left panel}) and by $\epsilon = 0.02$ (\emph{right panel}). The initial angles of the points in the green regions are fixed at $\omega_0 = 0, \Omega_0 = \pi$ and the ones of the points in the black regions are fixed at $\omega_0 = 0, \Omega_0 = 0$. In both panels, there are three distinct regions, denoted by I, II and III from the left to right. When the eccentricities are greater than a critical value $e_c(\epsilon)$ (the location of $e_c(\epsilon)$ is marked by a vertical red dashed line), orbit flips can take place from nearly coplanar configurations. Comparing the flipping regions shown in the left and right panels, we can see that the flipping area is larger when the factor $\epsilon$ is higher.}
\label{Fig1}
\end{figure*}

The numerical trajectories are propagated under the dynamical model specified by $\epsilon$ over $1.0 \times 10^3$ normalized unit of time with the initial condition $(e_0,i_0,\Omega_0,\omega_0)$. The numerical trajectory is identified as a flipping orbit once its inclination switches between prograde and retrograde. The results under the dynamical models characterized by $\epsilon = 0.01$ and $\epsilon = 0.02$ are reported in Fig. \ref{Fig1}, where the initial states with flips are distributed in the $(e_0,i_0)$ space\footnote{The initial angles $\Omega_0$ and $\omega_0$ are given. Thus, the full information of states for the points shown in Fig. \ref{Fig1} is known and each point in the $(e_0,i_0)$ space stands for a specific flipping orbit.}. In particular, the regions corresponding to case I are shown in green and the regions corresponding to case II are shown in black.

From Fig. \ref{Fig1}, we can observe that (a) there are three distinct regions of orbit flips shown in the $(e_0,i_0)$ space, denoted by I, II and III from the left to right, (b) the regions of orbit flips are symmetric with respect to $i_0 = 90^{\circ}$ due to the symmetry of the Hamiltonian under the octupole-level approximation, (c) the points with flips in case I are distributed in the low-eccentricity region (region I) and in the high-eccentricity region (region III), while the points with flips in case II are distributed in the intermediate-eccentricity region (region II), and (d) there exists a critical value of eccentricity $e_c$ in region III shown by red dashed lines and the orbits can flip from nearly-coplanar configurations when the eccentricity is greater than the critical value. Under dynamical models with different $\epsilon$, we can see that the structures of flipping regions are qualitatively similar. Quantitatively speaking, the regions of orbit flips are larger when the factor of the dynamical model $\epsilon$ is higher. Unless otherwise stated, in following simulations we mainly take the dynamical model specified by $\epsilon = 0.01$ as examples\footnote{Without doubt, the method adopted in this work is applicable to dynamical models with different values of $\epsilon$.}.

According to the magnitude of initial eccentricity and inclination, \citet{li2014eccentricity} classified the orbit flips as the high-eccentricity low-inclination case (called HeLi) and the low-eccentricity high-inclination case (LeHi). Obviously, the HeLi case corresponds to region I and LeHi case corresponds to region III in Fig. \ref{Fig1}. In their studies, orbit flips in the intermediate-eccentricity region are absent (see Fig. 10 in their work). Regarding the difference of underlying mechanism, the HeLi case is dominated by only the octupole order resonances, while the LeHi case is dominated by both the quadrature order resonances and the octupole order resonances \citep{li2014eccentricity, li2014}. However, it is still not clear about which resonance dominates orbit flips and how it works. About this topic, a recent advance was made by \citet{sidorenko2018eccentric}, who demonstrated that the orbit flips caused by the eccentric von Zeipel--Lidov--Kozai effect in the low-eccentricity high-inclination regions (region I and part of region II shown in Fig. \ref{Fig1}) can be interpreted as a resonance phenomenon with the critical argument $\sigma = h + {\rm sign}{(\cos{i})} g$ librating around zero or $\pi$. This interpretation is essential for us to deeply understand the mechanism of orbit flips.

In the following three sections, we intend to explore the numerical structures of flipping regions arising in the $(e_0,i_0)$ space from three different aspects: (a) Poincar\'e surfaces of section in Sect. \ref{Sect4}, (b) dynamical system theory (periodic orbits and invariant manifolds) in Sect. \ref{Sect5} and (c) perturbative treatments in Sect. \ref{Sect6}. Using each approach, it is possible to produce the boundaries of orbit flips.

\section{Surfaces of section}
\label{Sect4}

The dynamical model represented by equation (\ref{Eq2}) is of two degree of freedom but with one conserved quantity (the Hamiltonian), thus the dynamical model is not integral \citep{lithwick2011eccentric}. For such a kind of non-integrable systems, Poincar\'e surface of section provides a powerful tool to numerically explore the global structures in the phase space. The technique of Poincar\'e sections has been used in different contexts \citep{lithwick2011eccentric, li2014, naoz2017eccentric, malhotra2020divergence, lei2021dynamical, lei2021structures}. In this section, our purpose is to analyse the surfaces of section at different levels of Hamiltonian to find libration islands causing orbital flips.

In practice, the Poincar\'e surface of section is defined by
\begin{equation}\label{Eq4}
g = 0,\quad \dot g > 0.
\end{equation}
This definition of Poincar\'e sections has been adopted by \citet{li2014} and \citet{lithwick2011eccentric} in their studies. In numerical simulations, all the points with $g = 0$ and $\dot g > 0$ are recorded and presented in the $(h,H)$ space. At each point on the sections, $g$ is known (as zero) and the variable $G$ can be solved from the given Hamiltonian ${\cal H}(g=0,G,h,H)$. On the Poincar\'e sections, the smooth curves represent regular orbits, including librating trajectories inside the islands and circulating trajectories outside the islands, and the scattering points filling with a certain region on the sections stand for chaotic motions \citep{lithwick2011eccentric, li2014, lei2021structures}.

\begin{figure*}
\centering
\includegraphics[width=0.45\textwidth]{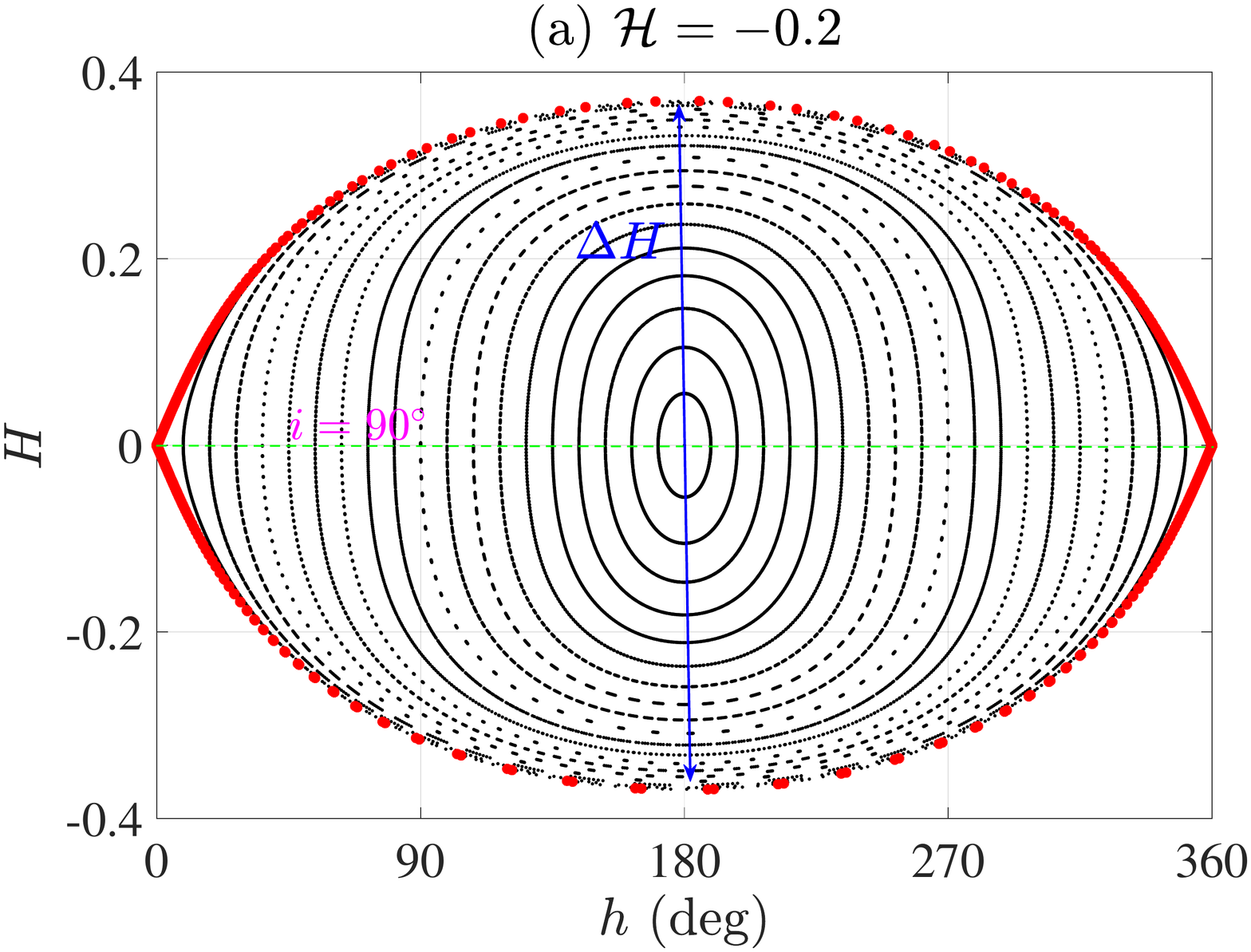}
\includegraphics[width=0.45\textwidth]{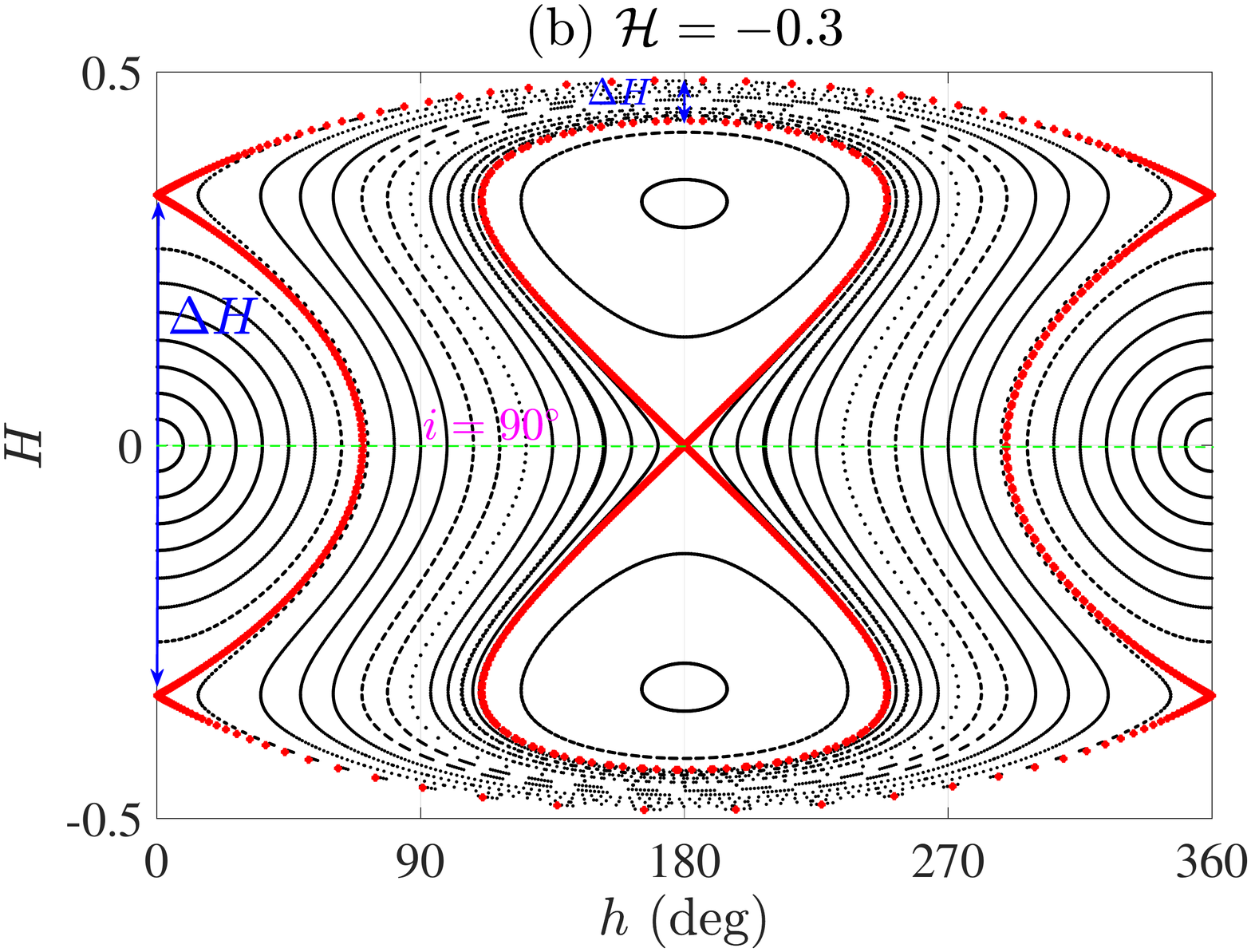}\\
\includegraphics[width=0.45\textwidth]{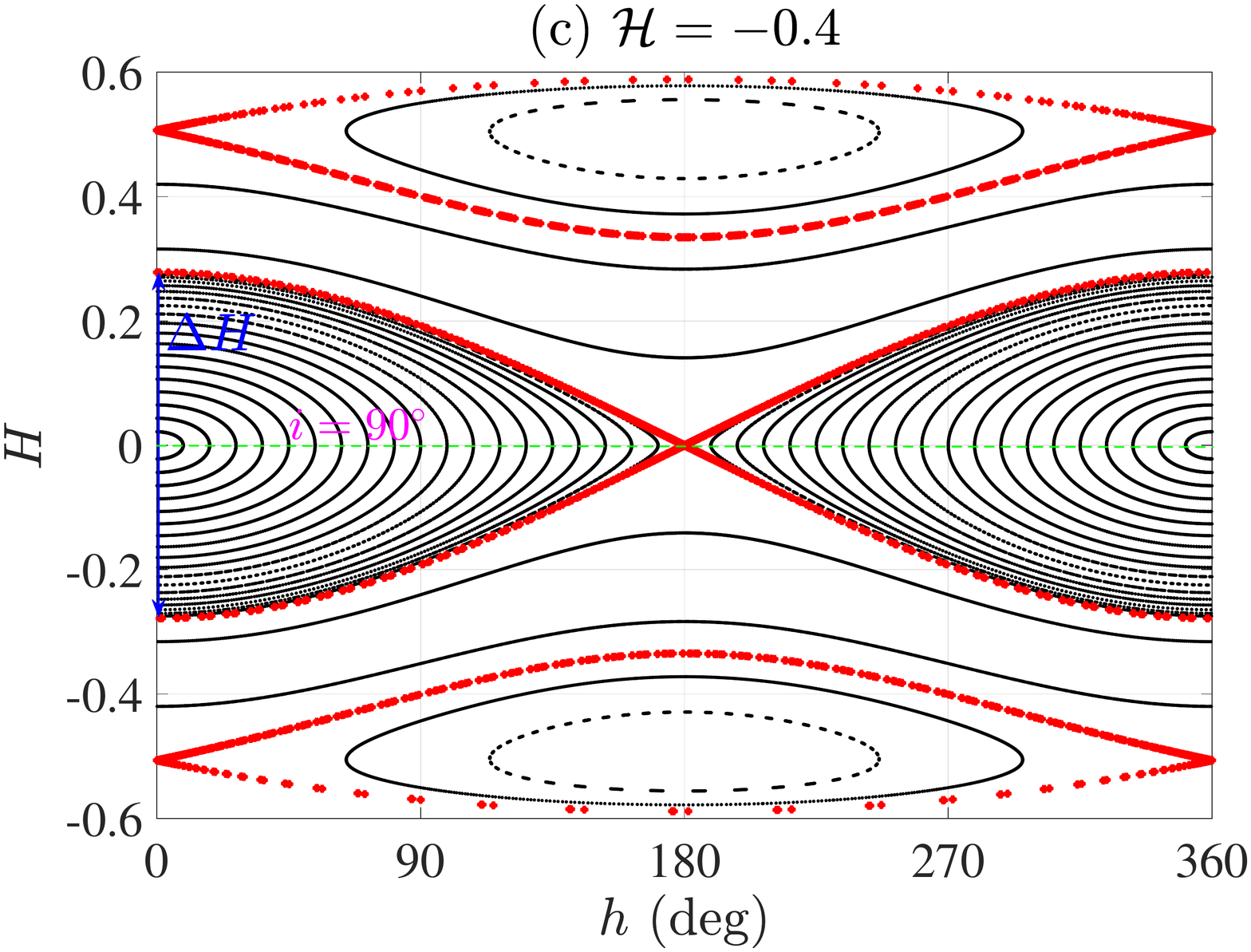}
\includegraphics[width=0.45\textwidth]{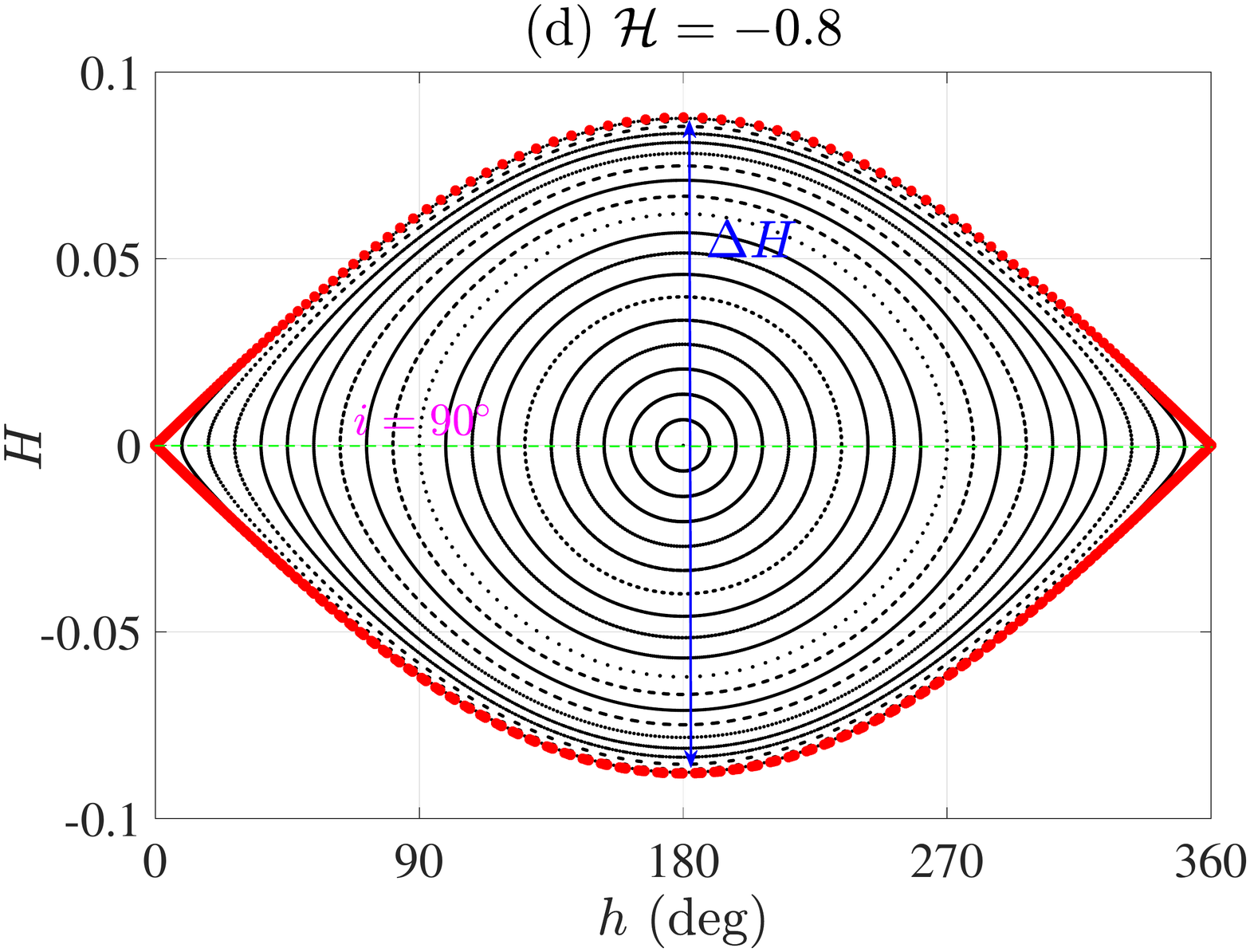}
\caption{Poincar\'e surfaces of section (defined by $g = 0$ and $\dot g > 0$) at different levels of Hamiltonian shown in the $(h,H)$ plane. In each panel, the green dashed line stands for the location of $i=90^{\circ}$ ($H = 0$) and $\Delta H$ represents the size of the flip region. The red dots stand for the boundaries of libration islands arising in the sections.}
\label{Fig3}
\end{figure*}

In Fig. \ref{Fig3}, the Poincar\'e surfaces of section are reported in the $(h,H)$ plane for Hamiltonian at ${\cal H} = -0.2$, ${\cal H} = -0.3$, ${\cal H} = -0.4$ and ${\cal H} = -0.8$. In each panel, the location of $i=90^{\circ}$ is marked in green dashed line and the dynamical separatrices are shown in red dots. The structures are different at different levels of Hamiltonian.

When the Hamiltonian is at ${\cal H} = -0.2$ (see the upper-left panel of Fig. \ref{Fig3}), there is a single libration island centered at $h=180^{\circ}$. We can see that (a) the saddle points are located at $h=0$, (b) both the saddle and elliptic points are located at $H=0$ ($i=90^{\circ}$), (c) almost all trajectories inside the island are regular, showing that the chaotic layer around the separatrix is very narrow and can be neglected\footnote{The chaotic layer increases with the perturbation and thus it will be thicker with larger $\epsilon$.}, and (d) all the trajectories inside the island of libration can realize flips between prograde and retrograde.

When the Hamiltonian is decreased to ${\cal H} = -0.3$ (see the upper-right panel of Fig. \ref{Fig3}), the dynamical structure arising in the Poincar\'e section becomes complicated. The structures are symmetric with respect to $H=0$, which is ensured by the symmetry of the Hamiltonian. Totally, there are six equilibrium points in the Poincar\'e section: two are located at $H=0$ and the remaining four are located at $H \ne 0$. There are three islands: one is around the centre $(h=0,H=0)$ and the other two centered at $(h=0,H \ne 0)$ are symmetric with respect to the polar line. All the trajectories inside the island centered at $(h=0,H=0)$ can flip between prograde and retrograde. In addition, it is observed that the trajectories between the inner and outer boundaries (shown by red dots) can also flip and these flipping trajectories hold the center at $(h=\pi,H=0)$.

When the Hamiltonian is further decreased to ${\cal H} = -0.4$ (see the bottom-left panel of Fig. \ref{Fig3}), there are three islands in the Poincar\'e sections. One island is around $(h=0,H=0)$ and the other two are symmetric with respect to $H=0$ and their centers are at $(h=\pi, H \ne 0)$. The trajectories inside the island around the symmetric center $(h=0,H=0)$ can realize flips. Those trajectories inside the islands around asymmetric centres can not flip.

Finally, when the Hamiltonian is taken as ${\cal H} = -0.8$ (see the bottom-right panel of Fig. \ref{Fig3}), the dynamical structure becomes simple and it is similar to the upper-left panel of Fig. \ref{Fig3}. There is a single island which is around the center $(h=\pi, H = 0)$. The trajectories inside the island are regular and all of them can flip between prograde and retrograde.

\begin{figure*}
\centering
\includegraphics[width=0.6\textwidth]{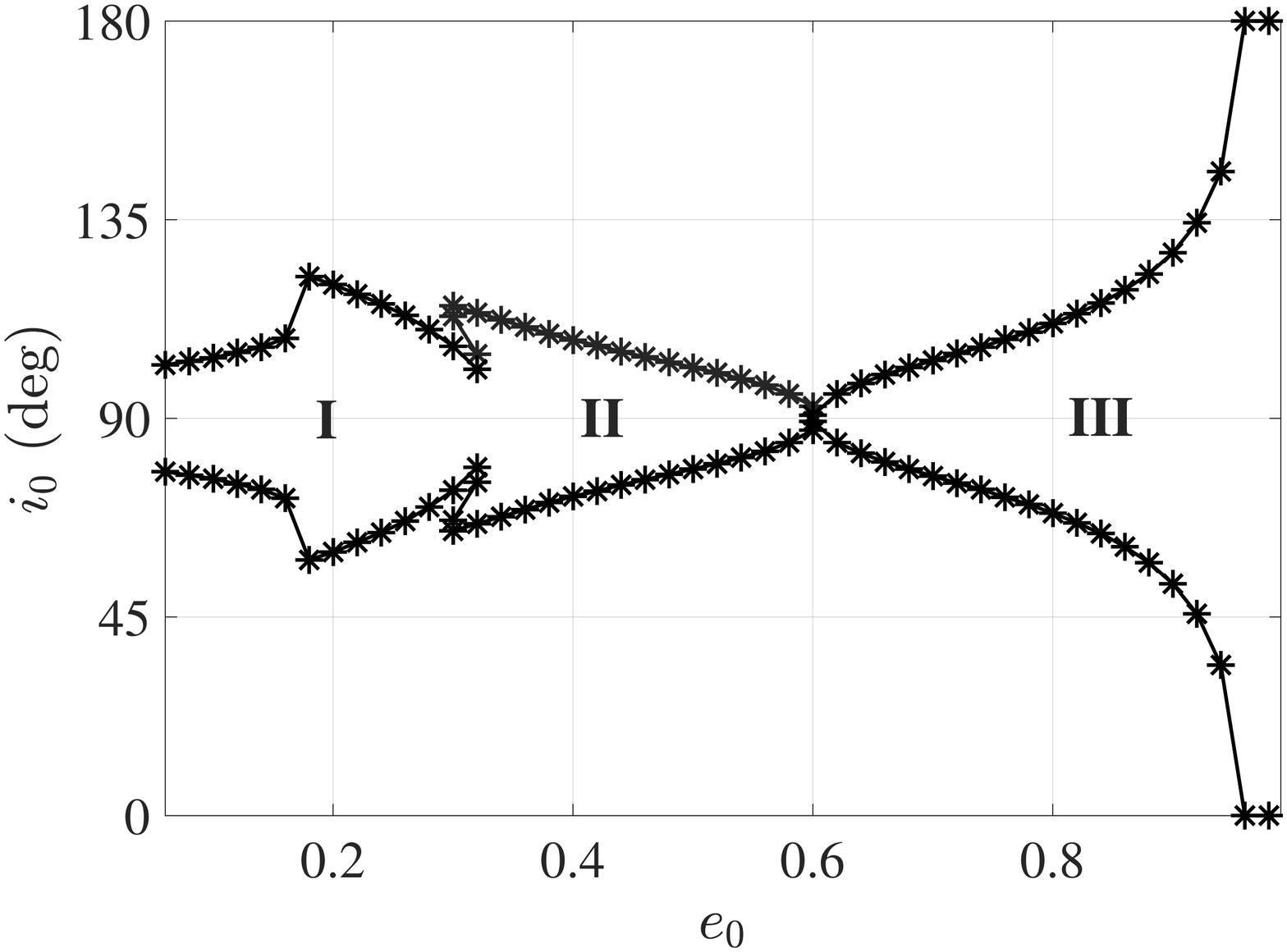}
\caption{Boundaries for the regions of orbit flips in the $(e_0, i_0)$ space obtained by analyzing the Poincar\'e surfaces of section (defined by $g = 0$ and $\dot g > 0$). The regions of orbit flips are denoted by I, II and III from the left to right. In particular, the longitude of ascending node is taken as $h=\pi$ in regions I and III and it is $h=0$ in region II.}
\label{Fig4}
\end{figure*}

In all panels of Fig. \ref{Fig3}, we use $\Delta H$ to stand for the size of libration islands causing orbit flips\footnote{$\Delta H$ represents the width of the libration island where orbits can flip and it is evaluated at the center of island.}. By analysing the Poincar\'e sections \citep{malhotra2020divergence, lei2021dynamical, lei2021structures}, it is possible to generate the size of flipping regions as a function of the Hamiltonian $\Delta H ({\cal H})$. According to the definition of $H$, we can produce the boundaries of flipping regions in the space $(e_0, i_0)$, as shown in Fig. \ref{Fig4}. From the left to right, the regions of orbit flips are denoted by I, II and III. In regions I and III, the boundaries (or $\Delta H$) are evaluated at $h=\pi$ (center of island) and they are evaluated at $h=0$ (center of island) in region II. Inside the regions between the boundaries, the orbits can flip between prograde and retrograde. It should be noted that our numerical boundaries shown in Fig. \ref{Fig4} are in agreement with the results shown by \citet{lithwick2011eccentric} for the case of $\epsilon = 0.01$ in the eccentricity range $e_0 \in [0,0.5]$ (see Fig. 8 in their work).

\section{Periodic orbits and invariant manifolds}
\label{Sect5}

In the previous section, we have produced the Poincar\'e sections and identified the islands of libration causing orbital flips. Dynamical system theory indicates us that, at the center of the libration island, there is a stable periodic orbit and the smooth curves inside the libration island correspond to quasi-periodic orbits \citep{winter2000stability, dutt2010analysis}. In addition, the saddle points arising in the sections correspond to unstable periodic orbits and the boundaries separating circulation regions from libration regions correspond to unstable/stable invariant manifolds emanating from the unstable periodic orbit with the same Hamiltonian level \citep{koon2000heteroclinic, koon2011Dynamical}. Motivated by dynamical system theory, we can explore the problem of orbit flips from the viewpoint of periodic orbits and invariant manifolds.

In particular, when $H=0$ and $h=0$ (or $h=\pi$), the Hamiltonian canonical equations read
\begin{equation}\label{Eq5}
\begin{aligned}
\dot h &= \frac{{\partial {\cal H}}}{{\partial H}} \equiv 0,\quad \dot H =  - \frac{{\partial {\cal H}}}{{\partial h}} \equiv 0,\\
\dot g &= \frac{{\partial {\cal H}}}{{\partial G}},\quad \dot G =  - \frac{{\partial {\cal H}}}{{\partial g}},
\end{aligned}
\end{equation}
which shows that $H$ and $h$ remain stationary. As a result, the dynamical model reduces to a single-degree-of freedom system under the condition of $H=0$ and $h=0$ (or $h=\pi$) and the solutions of the reduced dynamical model are periodic. Evidently, the periodic orbits correspond to the level curves of Hamiltonian in the $(g,G)$ space under the condition of $H=0$ and $h=0$ (or $h=\pi$). At a certain level of Hamiltonian, there are two periodic orbits corresponding to $h=0$ and $h=\pi$ and the analysis of linear stability shows that one is stable and the other one is unstable. Because of $H = 0$ (i.e., $i = 90^{\circ}$), these special periodic orbits are called polar periodic orbits.

\begin{figure*}
\centering
\includegraphics[width=0.45\textwidth]{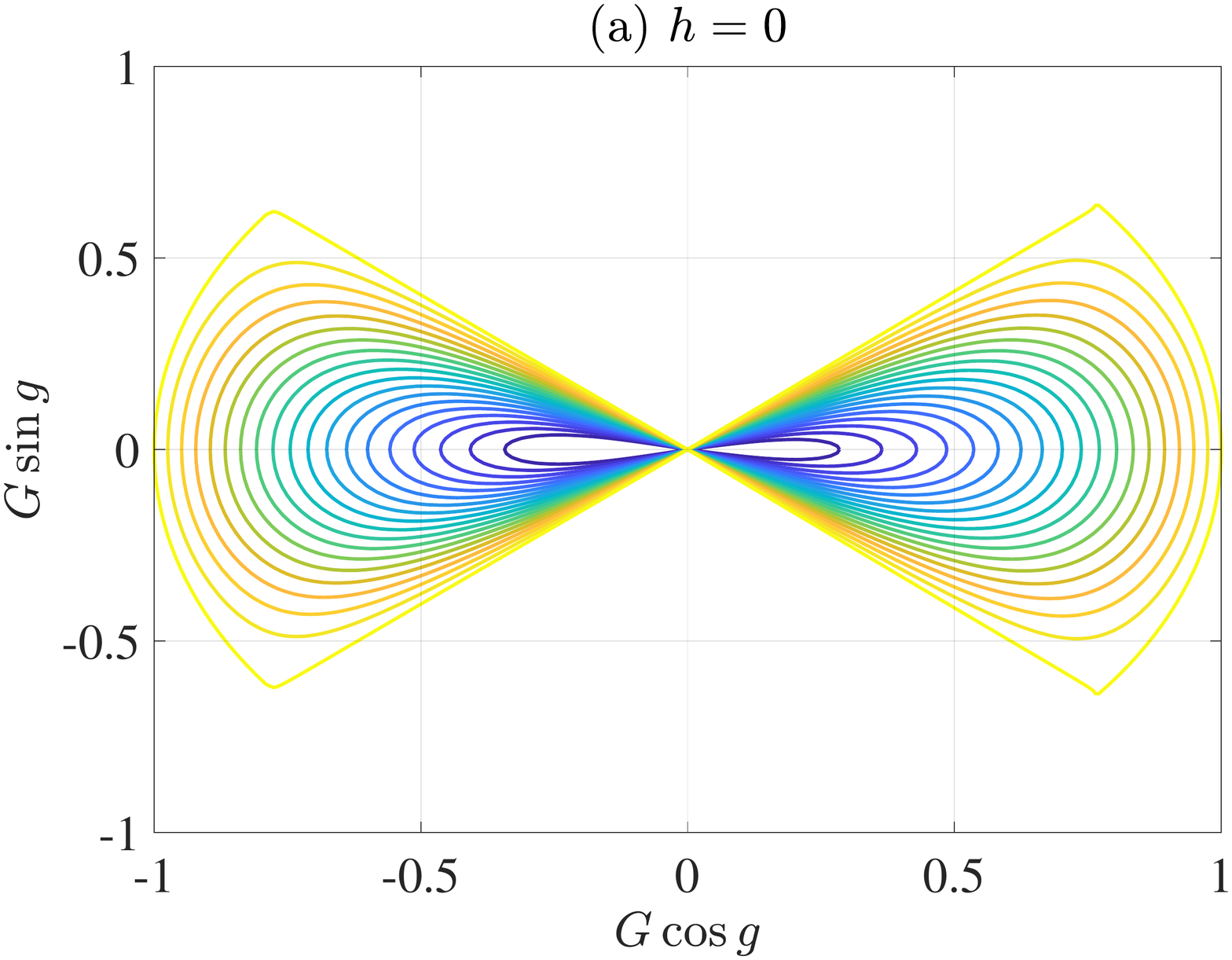}
\includegraphics[width=0.45\textwidth]{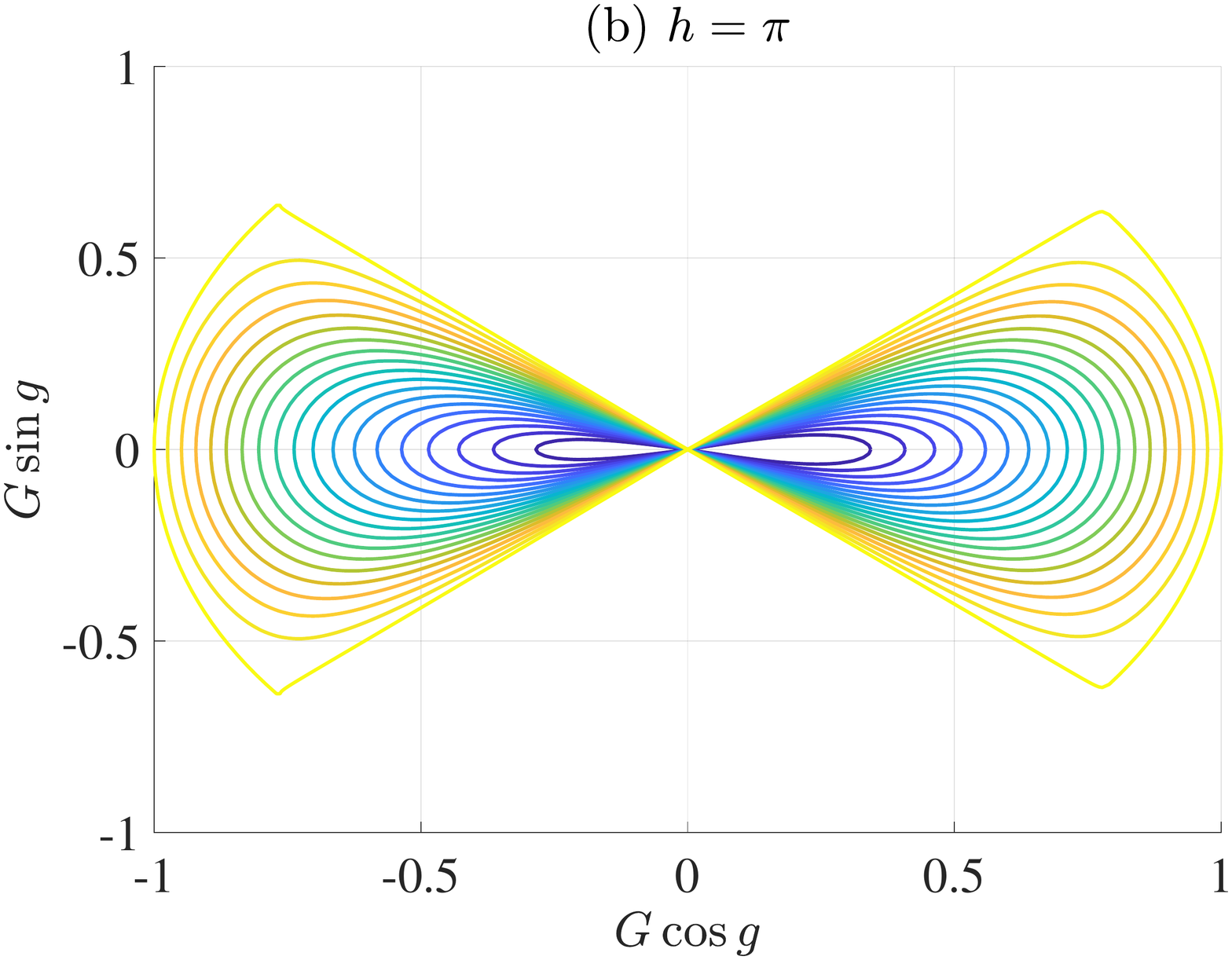}
\caption{Polar periodic orbits shown in polar coordinates ($G\cos{g},G\sin{g}$). The polar periodic orbits correspond to level curves of Hamiltonian under the condition of $H=0$ and $h=0$ (\emph{left panel}) and $H=0$ and $h=\pi$ (\emph{right panel}).}
\label{Fig5}
\end{figure*}

\begin{figure*}
\centering
\includegraphics[width=0.45\textwidth]{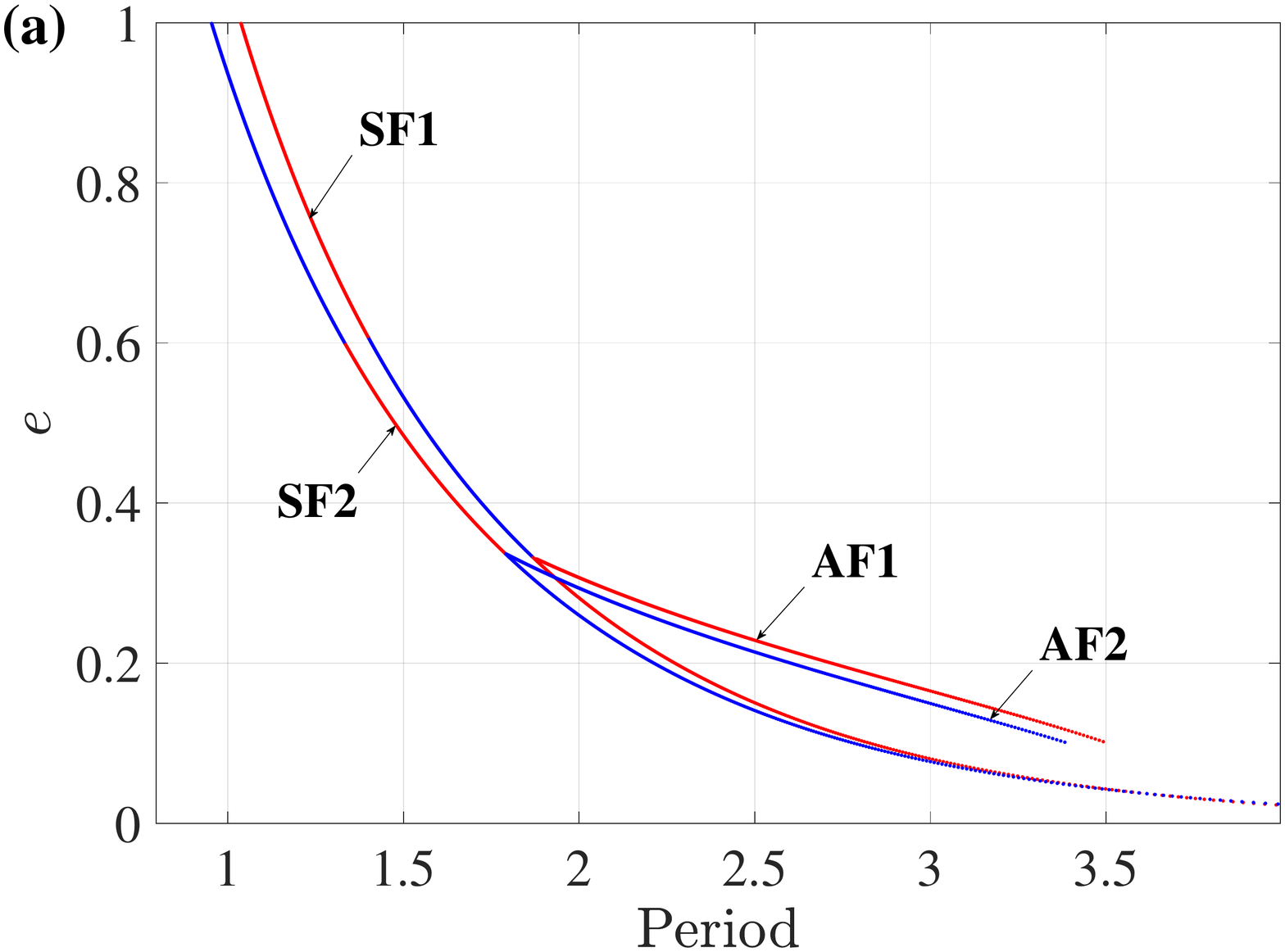}
\includegraphics[width=0.45\textwidth]{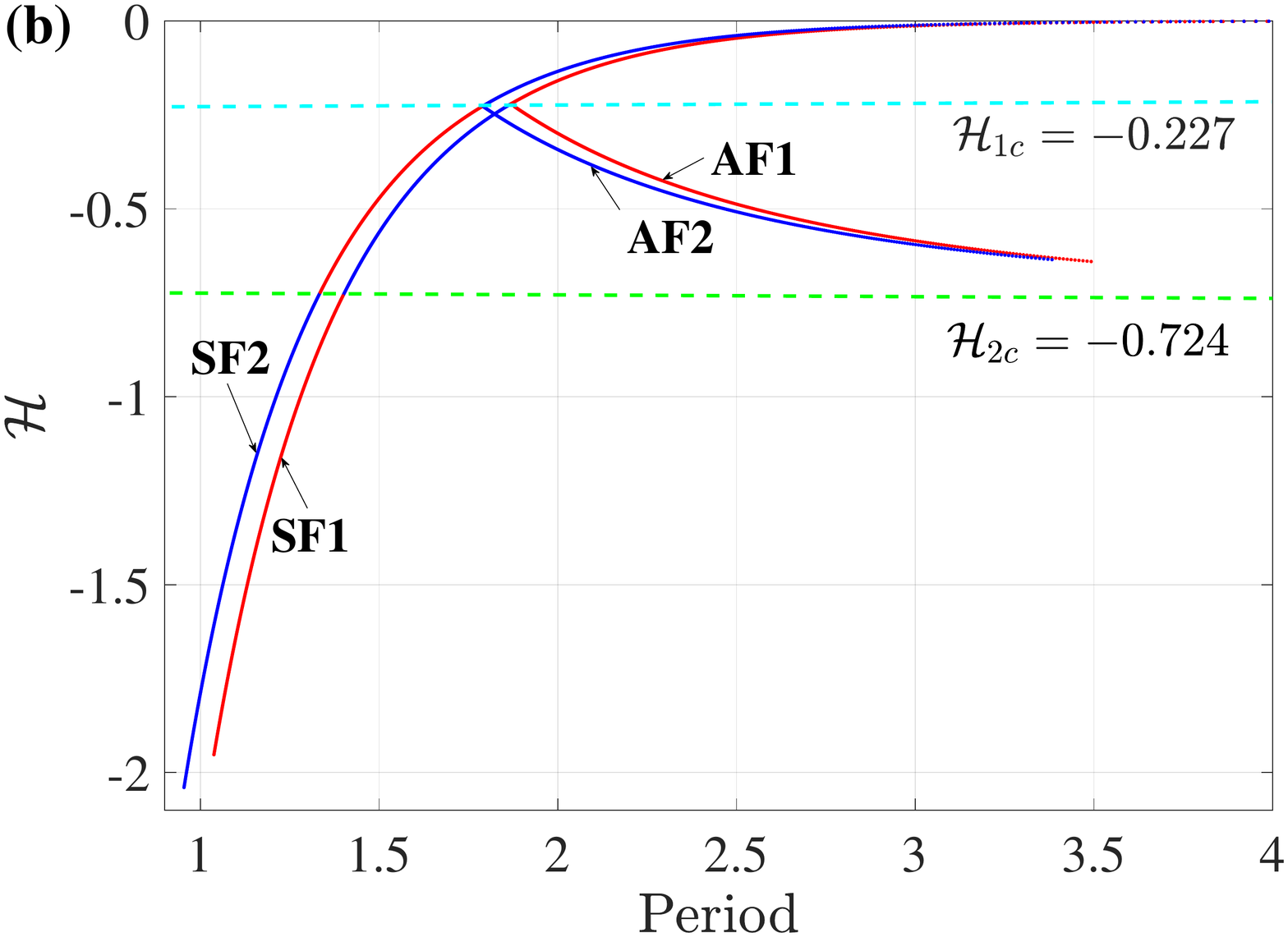}
\caption{Characteristic curves of polar periodic orbits in the plane of eccentricity and period (\emph{left panel}) and in the plane of Hamiltonian and period (\emph{right panel}). Blue lines stand for characteristic curves of stable periodic orbits, while red lines stand for characteristic curves of the unstable periodic orbits. `SF1' and `SF2' represent the polar families and `AF1' and `AF2' represent bifurcated families. The dashed lines marked in the right panel stand for the critical lines of Hamiltonian where the stability of periodic orbits changes.}
\label{Fig6}
\end{figure*}

The polar periodic orbits with Hamiltonian ranging from -1.8 to 0 with step of 0.1 are presented in Fig. \ref{Fig5}, where the left panel is for $H=0$ and $h=0$ and the right panel is for $H=0$ and $h=\pi$. In Fig. \ref{Fig6}, the characteristic curves of the polar periodic orbits are reported in the period--eccentricity plane (see the left panel) and in the period--Hamiltonian plane (see the right panel). The stability of polar periodic orbits is determined by the eigenvalues of the monodromy matrix, which is the state transition matrix evaluated at one period \citep{koon2011Dynamical}. The stable members are shown in blue and the unstable ones are shown in red. The families of polar periodic orbits are denoted by SF1 (corresponding to $h=0$) and SF2 (corresponding to $h=\pi$). In addition, two asymmetrical families of periodic orbits are bifurcated from the polar families and they are denoted by AF1 and AF2. The Hamiltonian of the bifurcation point is equal to ${\cal H}_{1c} = -0.227$.

For an unstable periodic orbit, the eigenvalues of the monodromy matrix are in the form of $(\lambda_1 =1, \lambda_2=1, \lambda_3, \lambda_4)$, where the stability index $s = \frac{1}{2} (\lambda_3 + \lambda_4)$ is greater than unity \citep{henon1969numerical, koon2011Dynamical}. Without loss of generality, we assume $|\lambda_3| >1$ and $|\lambda_4| < 1$. Thus, the eigenvector associated with $\lambda_3$ determines an asymptotically unstable direction and the one associated with $\lambda_4$ determines an asymptotically stable direction \citep{koon2011Dynamical}. Executing a small deviation along the unstable direction, the trajectory will depart from the periodic orbit exponentially when time tends to infinity. On the contrary, executing a small deviation along the stable direction, the trajectory will approach the periodic orbit exponentially when time tends to infinity (or it will depart from the periodic orbit when time tends to minus infinity). All asymptotically unstable and stable trajectories stemming from a certain unstable periodic orbit form a set of trajectories in the phase space, which is called invariant manifolds \citep{koon2011Dynamical}. Thus, the trajectories in the invariant manifolds share the same Hamiltonian with their host periodic orbit.

\begin{figure*}
\centering
\includegraphics[width=0.45\textwidth]{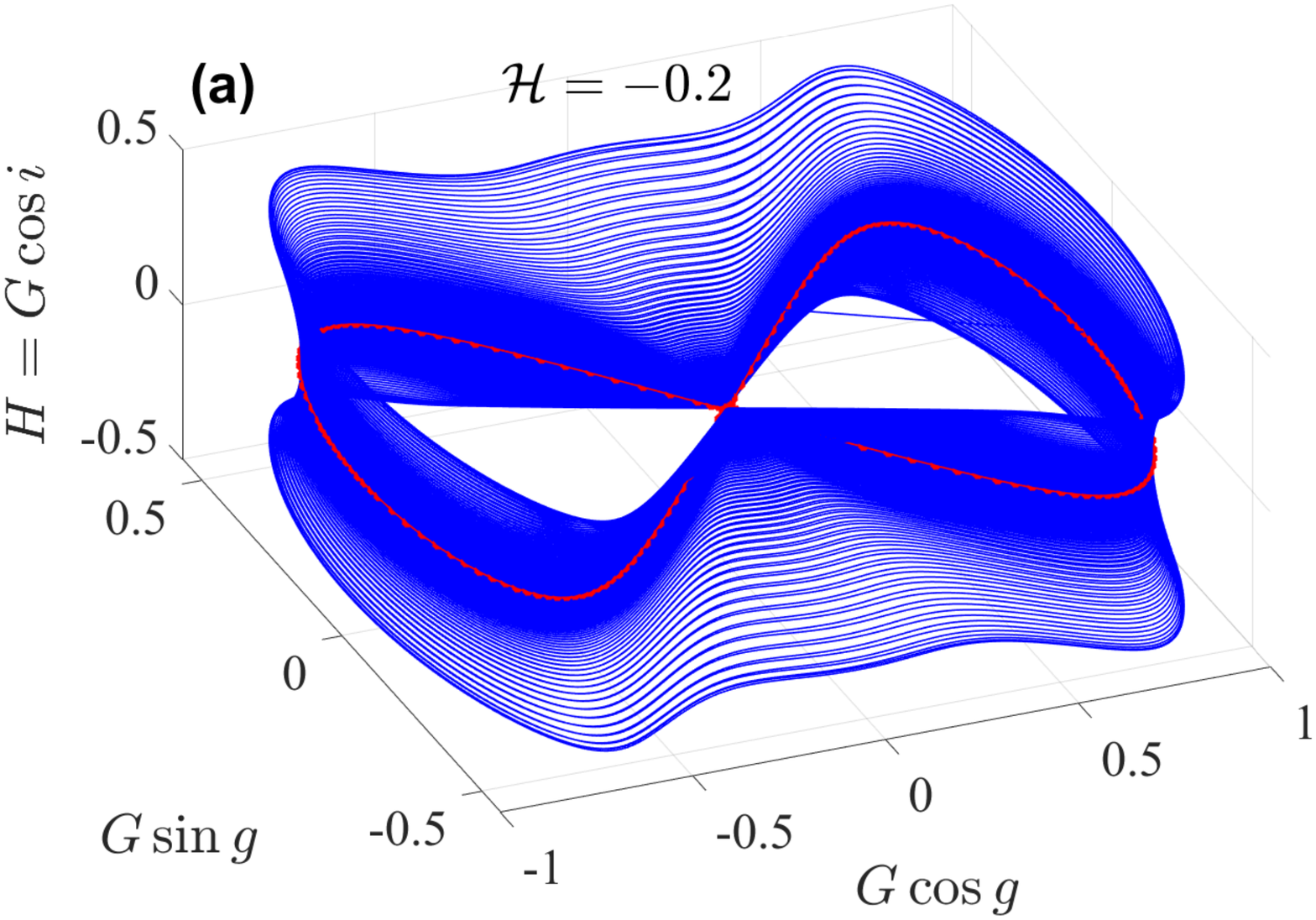}
\includegraphics[width=0.45\textwidth]{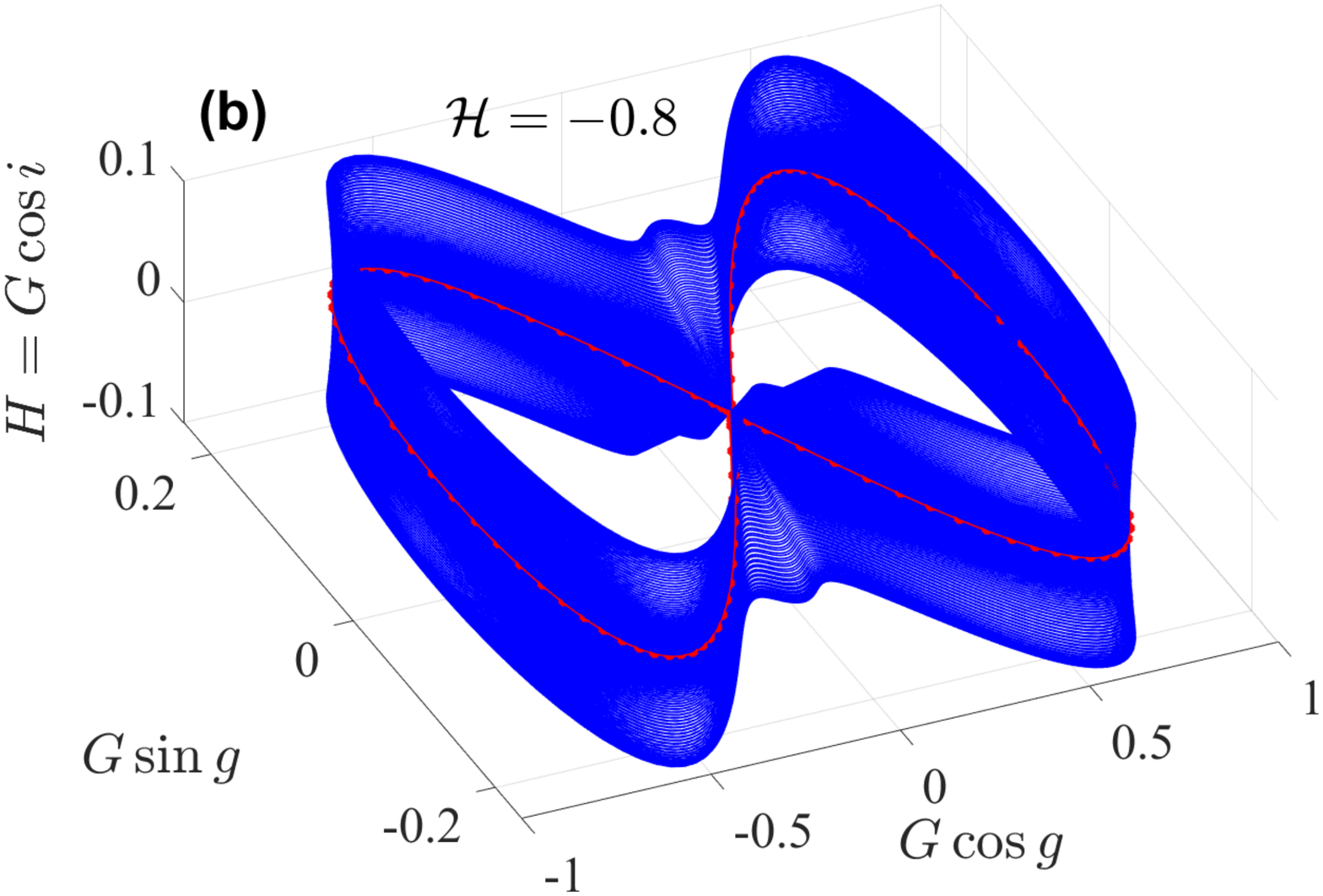}
\caption{Invariant manifolds (a set of asymptotically unstable and/or stable trajectories) stemming from the unstable periodic orbits with ${\cal H} = -0.20$ (\emph{left panel}) and ${\cal H} = -0.80$ (\emph{right panel}). The red lines stand for unstable polar orbits.}
\label{Fig7}
\end{figure*}

In Fig. \ref{Fig7}, the invariant manifolds emanating from the unstable periodic orbits with ${\cal H} = -0.2$ and ${\cal H} = -0.8$ are presented in the three-dimensional space $(G\cos{g},G\sin{g},H)$. The fourth-dimensional information $h$ can be obtained by the given level of Hamiltonian ${\cal H} (g,G,h,H)$. The unstable periodic orbits are shown in red lines. From Fig. \ref{Fig7}, it is observed that following along the trajectories in the invariant manifolds the orbital inclination can change from prograde to polar state or from retrograde to polar state. Evidently, the invariant manifolds provide a high-dimensional invariant surface embedded in the phase space. To see the dynamical role played by the manifolds, we also take advantage of the tool of Poincar\'e sections to project invariant manifolds onto a two-dimensional plane \citep{koon2000heteroclinic, koon2001low}.

\begin{figure*}
\centering
\includegraphics[width=0.45\textwidth]{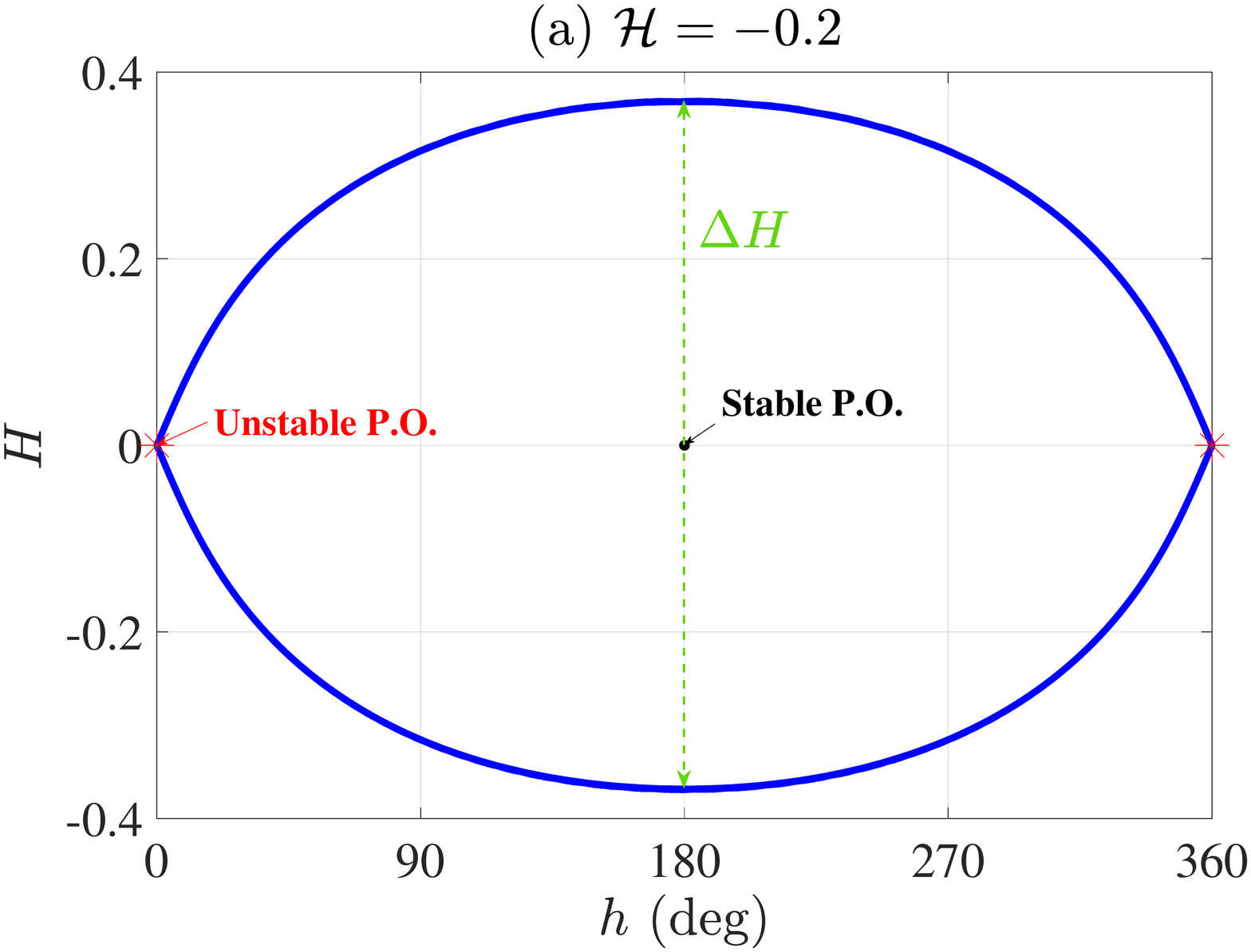}
\includegraphics[width=0.45\textwidth]{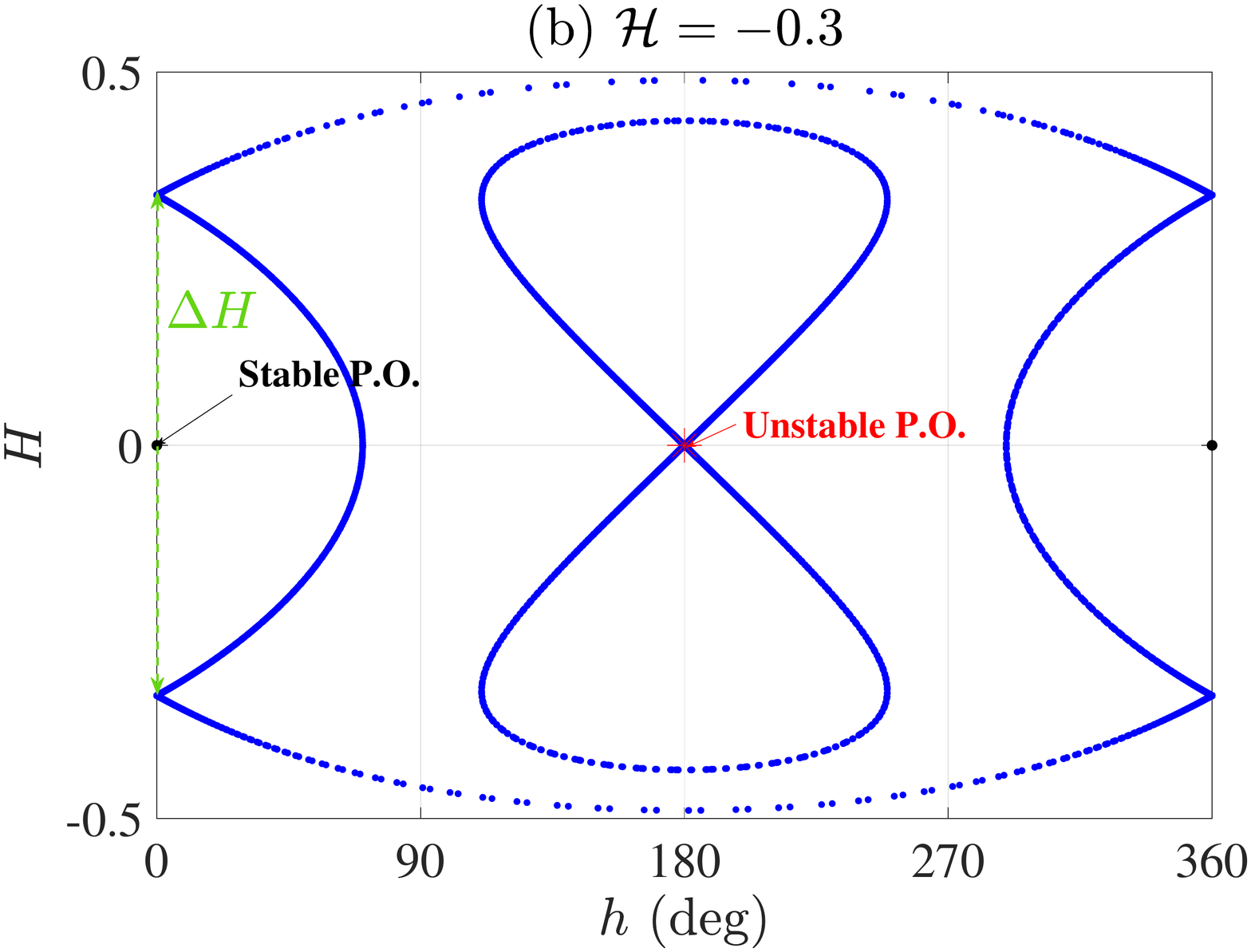}\\
\includegraphics[width=0.45\textwidth]{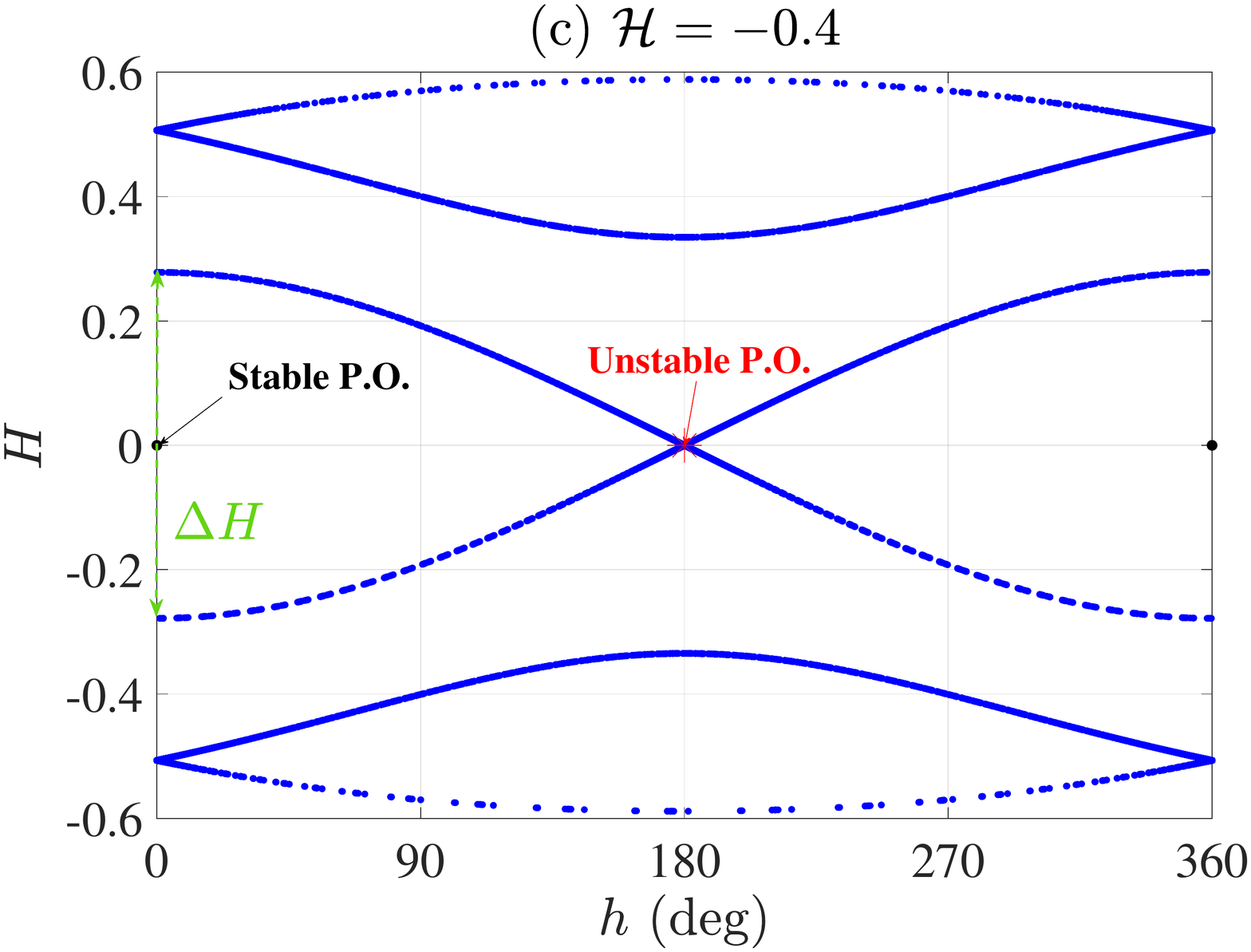}
\includegraphics[width=0.45\textwidth]{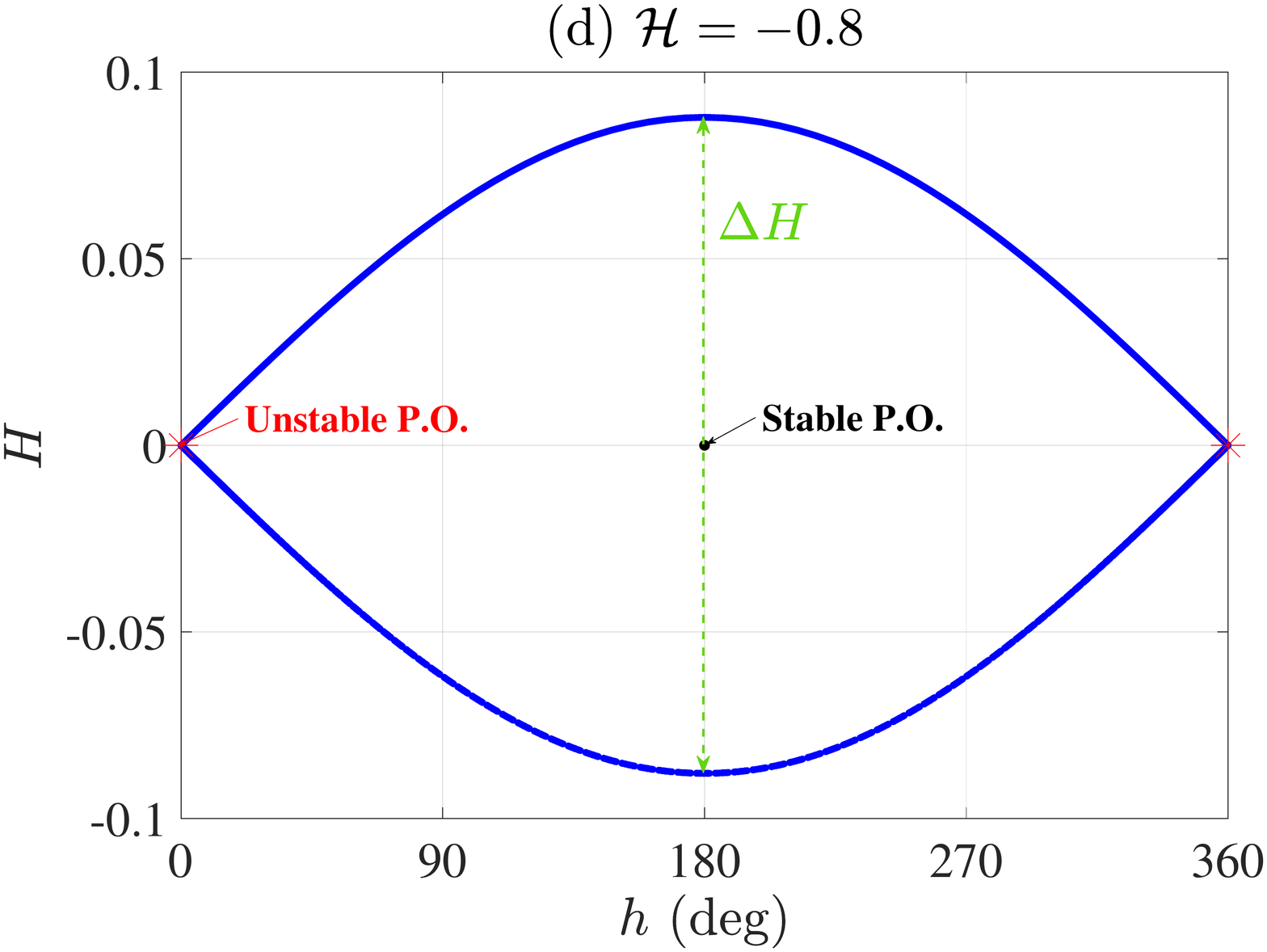}
\caption{Poincar\'e surfaces of section (defined by $g=0$ and $\dot g >0$) for the invariant manifolds associated with unstable polar periodic orbits at different levels of Hamiltonian. The locations of the stable and unstable polar periodic orbits are marked. It is evident that the sections of invariant manifolds provide boundaries of libration islands in the phase space.}
\label{Fig8}
\end{figure*}

In Fig. \ref{Fig8}, the Poincar\'e sections (defined by $g=0$) of the invariant manifolds with different levels of Hamiltonian are presented in the $(h,H)$ space. For convenience, the locations of the stable and unstable (polar) periodic orbits are marked. Comparing Fig. \ref{Fig8} with Fig. \ref{Fig3}, it is evident that the sections of invariant manifolds provided boundaries for the libration islands centered at the location of stable periodic orbit, i.e., providing boundaries for those quasi-periodic orbits \citep{koon2011Dynamical}. Certainly, those islands of libration causing orbital flips are bounded by the sections of invariant manifolds. Therefore, it is possible to produce the boundaries of orbit flips by analysing the invariant manifolds with different levels of Hamiltonian.

\begin{figure*}
\centering
\includegraphics[width=0.6\textwidth]{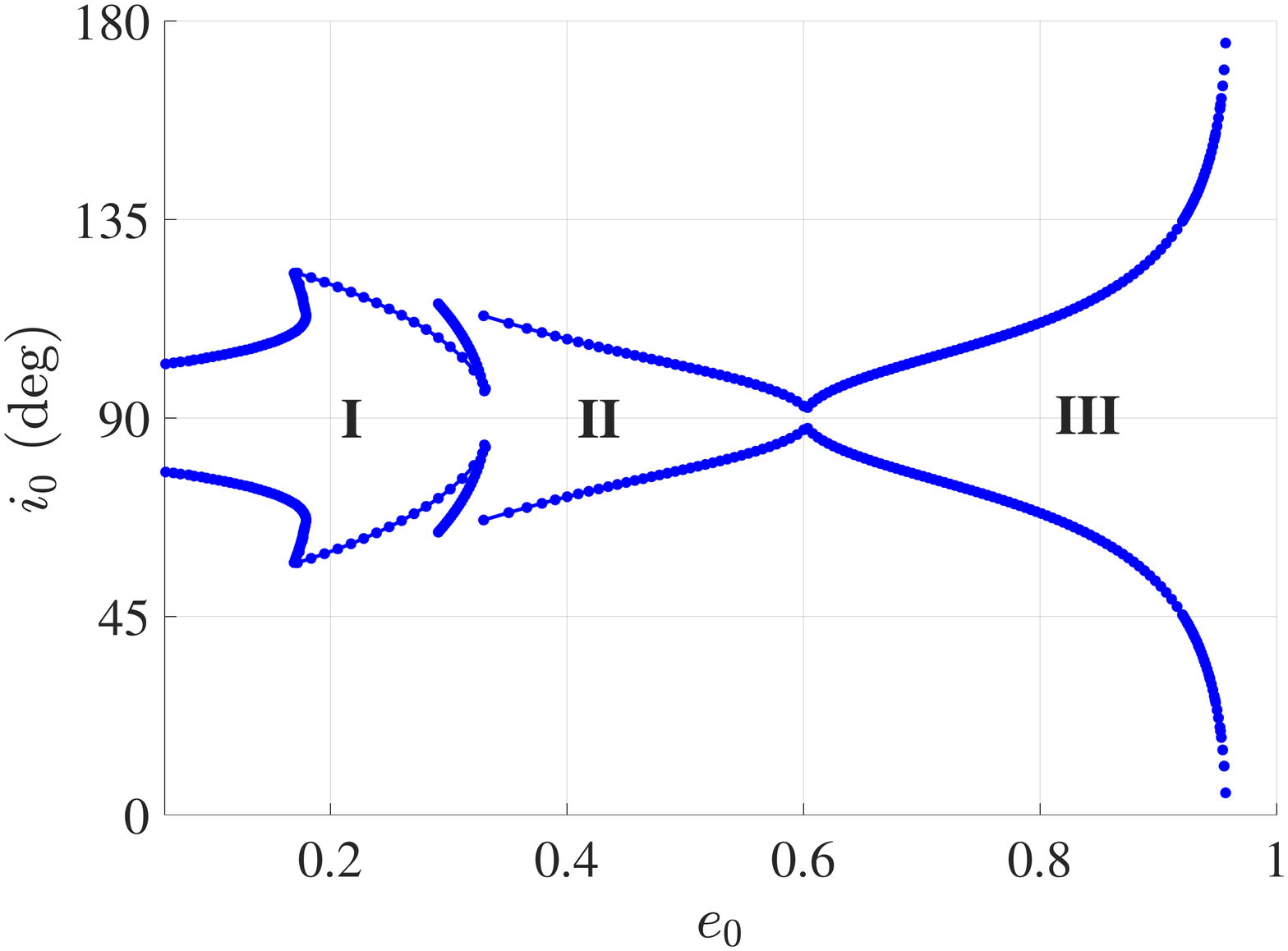}
\caption{Boundaries of the regions of orbit flips in the $(e_0, i_0)$ space obtained by analysing invariant manifolds associated with unstable polar orbits. For convenience, the regions of orbit flips are denoted by I, II and III from the left to right. In regions I and III, the longitude of ascending node is taken as $h=\pi$ and, in region II, it is taken as $h=0$.}
\label{Fig9}
\end{figure*}

By analysing the sections (defined by $g=0$) of invariant manifolds, the boundaries of orbit flips are shown in the $(e_0,i_0)$ space, as reported in Fig. \ref{Fig9}. Similarly, there are three distinct regions of orbital flips, denoted by I, II and III from the left to right. In regions I and III, the boundaries are evaluated at $h=\pi$ (location of stable periodic orbit) and, in region II, the boundaries are evaluated at $h=0$ (location of stable periodic orbit). Inside the regions enclosed the boundaries provided by invariant manifolds, the points correspond to quasi-periodic orbits around polar periodic orbits and these quasi-periodic orbits can realize flips between prograde and retrograde. Therefore, we can understand that the essence of flipping orbits is a kind of quasi-periodic orbits around the polar orbits and naturally the quasi-periodic orbits (flipping orbits) are bounded by invariant manifolds with the same Hamiltonian.

\section{Perturbative treatments}
\label{Sect6}

Orbit flips caused by the eccentric von Zeipel--Lidov--Kozai effect \citep{naoz2011hot,li2014,li2014eccentricity,katz2011long} are interpreted as a resonance phenomenon where the critical argument $\sigma = h + {\rm sign}(H) g$ is librating around zero or $\pi$ \citep{sidorenko2018eccentric}. Based on such a viewpoint, in this section we further study the problem of orbit flips by taking advantage of perturbation treatments \citep{wisdom1985perturbative, neishtadt1987change, neishtadt2004wisdom, sidorenko2014quasi, saillenfest2016long, efimov2020analytically}.

\begin{figure}
\centering
\includegraphics[width=0.48\textwidth]{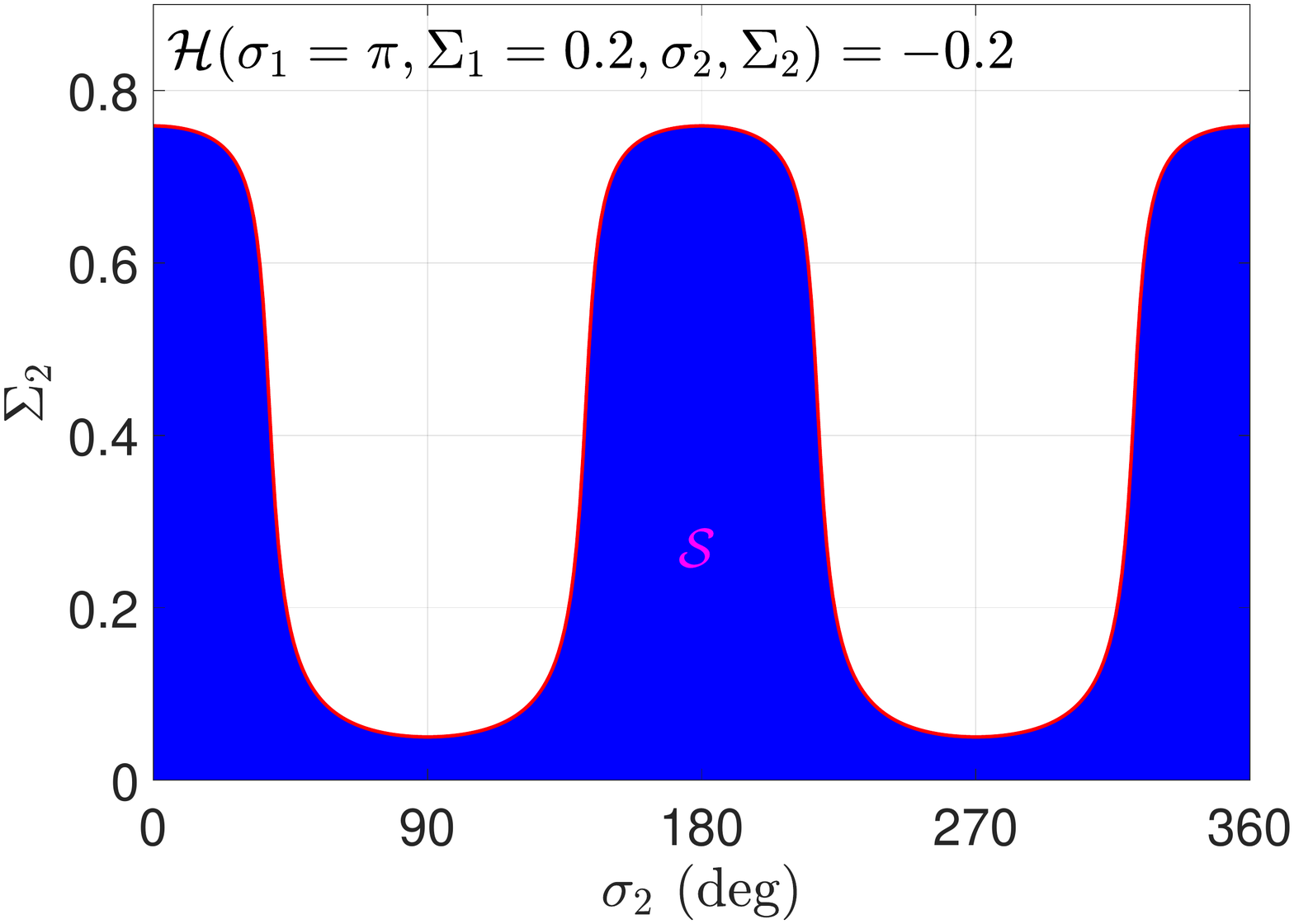}
\caption{The solution curve (red line) of ${\cal H} (\sigma_1=\pi,\Sigma_1=0.2,\sigma_2, \Sigma_2) = -0.2$ shown in the $(\sigma_2, \Sigma_2)$ plane. The path integral of the solution curve $\Sigma_2 (\sigma_2)$ is defined as the adiabatic invariant ${\cal S}$, which corresponds to the area of the blue shaded region.}
\label{Fig9_1}
\end{figure}

To study the resonance associated with orbit flips, a new set of variables is introduced as follows \citep{sidorenko2018eccentric}:
\begin{equation}\label{Eq6}
\begin{aligned}
{\sigma _1} &= h + {\rm sign}(H)g,\quad {\Sigma _1} = H,\\
{\sigma _2} &= g,\quad {\Sigma _2} = G - \left| H \right|,
\end{aligned}
\end{equation}
which is a canonical transformation with the following generating function
\begin{equation}
{\cal S} = h{\Sigma _1} + g\left( {\left| {{\Sigma _1}} \right| + {\Sigma _2}} \right).
\end{equation}
Under such a set of variables, the Hamiltonian can be written as
\begin{equation}\label{Eq7}
\begin{aligned}
{\cal H}\left( {{\sigma _1},{\sigma _2},{\Sigma _1},{\Sigma _2}} \right) = & - {F_{\rm quad}}\left( {{\sigma _2},{\Sigma _1},{\Sigma _2}} \right) \\
&- \varepsilon {F_{\rm oct}}\left( {{\sigma _1},{\sigma _2},{\Sigma _1},{\Sigma _2}} \right)
\end{aligned}
\end{equation}
and the Hamiltonian canonical relations yield the equations of motion,
\begin{equation}
\begin{aligned}
{{\dot \sigma }_1} &= \frac{{\partial {\cal H}}}{{\partial {\Sigma _1}}},\quad {{\dot \Sigma }_1} =  - \frac{{\partial {\cal H}}}{{\partial {\sigma _1}}},\\
{{\dot \sigma }_2} &= \frac{{\partial {\cal H}}}{{\partial {\Sigma _2}}},\quad {{\dot \Sigma }_2} =  - \frac{{\partial {\cal H}}}{{\partial {\sigma _2}}}.
\end{aligned}
\end{equation}

In particular, when the test particle is located inside the resonance associated with orbit flips, the resonant angle $\sigma_1$ is a long-period variable and the angular variable $\sigma_2$ is a short-period variable, showing that the dynamical model is a typical separable Hamiltonian system \citep{henrard1990semi}. Under the resonance condition, the dynamical model can be separated into the fast and slow degrees of freedom (fast and slow DOFs), as discussed by \citet{sidorenko2018eccentric}. When facing the evolution associated with the fast DOF, the variables associated with the slow DOF can be treated as parameters \citep{neishtadt1987change, wisdom1985perturbative}. Such an approximation ``fixing the slow variables by steps" has been applied to different contexts \citep{henrard1987perturbative, yokoyama1996simple, henrard1989motion, milani1998dynamics, sidorenko2014quasi, sidorenko2018dynamics, sidorenko2020perturbative, efimov2020analytically, saillenfest2016long, saillenfest2017study, saillenfest2020long}. During the timescale of the fast DOF, the variables $(\sigma_1,\Sigma_1)$ associated with the slow DOF are considered as parameters, thus the subsystem associated with the fast DOF becomes integrable. The solution curves are level curves of the Hamiltonian in the $(\sigma_2, \Sigma_2)$ space and certainly they are periodic trajectories \citep{saillenfest2016long}.

For the considered problem, \citet{katz2011long} introduce the longitude $\Omega_e$ to describe the very long-term behaviors caused by von Zeipel--Lidov--Kozai effects. Some discussions about the longitude $\Omega_e$ used by \citet{katz2011long} and the critical argument $\sigma_1$ adopted in this current work are presented in the Appendix.

For the subsystem of the fast DOF (with $\sigma_1$ and $\Sigma_1$ as parameters), the Arnold action-angle variables are introduced as follows \citep{morbidelli2002modern}:
\begin{equation}\label{Eq8}
\Sigma _2^* = \frac{1}{{2\pi }}\int\limits_0^{2\pi } {{\Sigma _2}{\rm d}{\sigma _2}} ,\quad \sigma _2^* = \frac{{2\pi }}{T}t
\end{equation}
where $\Sigma_2^*$ corresponds to the path integration (divided by $2\pi$) of the solution curve $\Sigma_2 (\sigma_2)$ and $\sigma_2^*$ is a linear function of time ($T$ is the period of $\sigma_2$). The transformation given by Eq. (\ref{Eq8}) can be realized by the following generating function,
\begin{equation*}
\begin{aligned}
W\left( {{\sigma _1},{\Sigma _1};{\sigma _2},\Sigma _2^*} \right) &= \int\limits_0^{{\sigma _2}} {{\Sigma _2}\left( {{\cal H}\left( {\Sigma _2^*,{\sigma _1},{\Sigma _1}} \right),{\sigma _2}} \right){\rm d}{\sigma _2}} \\
&= \int\limits_0^t {{\Sigma _2}\frac{{\partial {\cal H}}}{{\partial {\Sigma _2}}}{\rm d}t}.
\end{aligned}
\end{equation*}

In terms of the action-angle variables, the Hamiltonian can be further written as
\begin{equation}\label{Eq9}
{\cal H}\left( {{\sigma _1},{\Sigma _1},\Sigma _2^*} \right) =  - {F_{\rm quad}}\left( {{\Sigma _1},{\Sigma _2^*}} \right) - \varepsilon {F_{\rm oct}}\left( {{\sigma _1},{\Sigma _1},{\Sigma _2^*}} \right)
\end{equation}
In the dynamical model represented by the Hamiltonian ${\cal H}\left( {{\sigma _1},{\Sigma _1},\Sigma _2^*} \right)$, the action variable $\Sigma_2^*$ becomes a motion integral \citep{neishtadt1987change}. We denote the motion integral $2\pi \Sigma_2^*$ as the adiabatic invariant in the long-term evolution \citep{henrard1993adiabatic}, which is defined by
\begin{equation}\label{Eq10}
\begin{aligned}
{\cal S}\left( {{\cal H},{\sigma _1},{\Sigma _1}} \right) \buildrel \Delta \over = & {2\pi} \Sigma _2^{\rm{*}}\left( {{\cal H},{\sigma _1},{\Sigma _1}} \right) \\
= & \int\limits_0^{2\pi} {{\Sigma _2}\left( {{\cal H},{\sigma _1},{\Sigma _1},{\sigma _2}} \right){\rm d}{\sigma _2}}.
\end{aligned}
\end{equation}
When the pair of parameters $(\sigma_1, \Sigma_1)$ and the Hamiltonian ${\cal H}$ are given and fixed, the solution curve $\Sigma_2 (\sigma_2)$ as well as the generating function can be produced by numerically integrating the following equations of motion over one period of $\sigma_2$,
\begin{equation*}
\left\{ \begin{aligned}
{{\dot \sigma }_2} &= \frac{{\partial {\cal H}}}{{\partial {\Sigma _2}}},\\
{{\dot \Sigma }_2} &=  - \frac{{\partial {\cal H}}}{{\partial {\sigma _2}}},\\
\dot W &= {\Sigma _2}\frac{{\partial {\cal H}}}{{\partial {\Sigma _2}}}.
\end{aligned} \right.
\end{equation*}
At the initial moment, $\sigma_2$ and $W$ are zero and the initial value of $\Sigma_2$ can be produced from the given Hamiltonian. When the time $t$ is equal to one period $T$, the angle $\sigma_2(T)$ is equal to $2\pi$ and the value of $W(T)$ corresponds to the defined adiabatic invariant ${\cal S}$. Evidently, the adiabatic invariant ${\cal S}$ is a function of the Hamiltonian ${\cal H}$ and the variables associated with the slow DOF $(\sigma_1, \Sigma_1)$.

The geometrical definition of the adiabatic invariant ${\cal S}({\cal H}, \sigma_1, \Sigma_1)$ is provided in Fig. \ref{Fig9_1}, where the adiabatic invariant ${\cal S}$ is equal to the signed area of the shaded region and the red curve stands for the solution curve $\Sigma_2 (\sigma_2)$ when the Hamiltonian is taken as ${\cal H} = -0.2$ and the variables associated with the slow DOF are taken as $\sigma_1 = \pi$ and $\Sigma_1 = 0.2$.

\begin{figure*}
\centering
\includegraphics[width=0.45\textwidth]{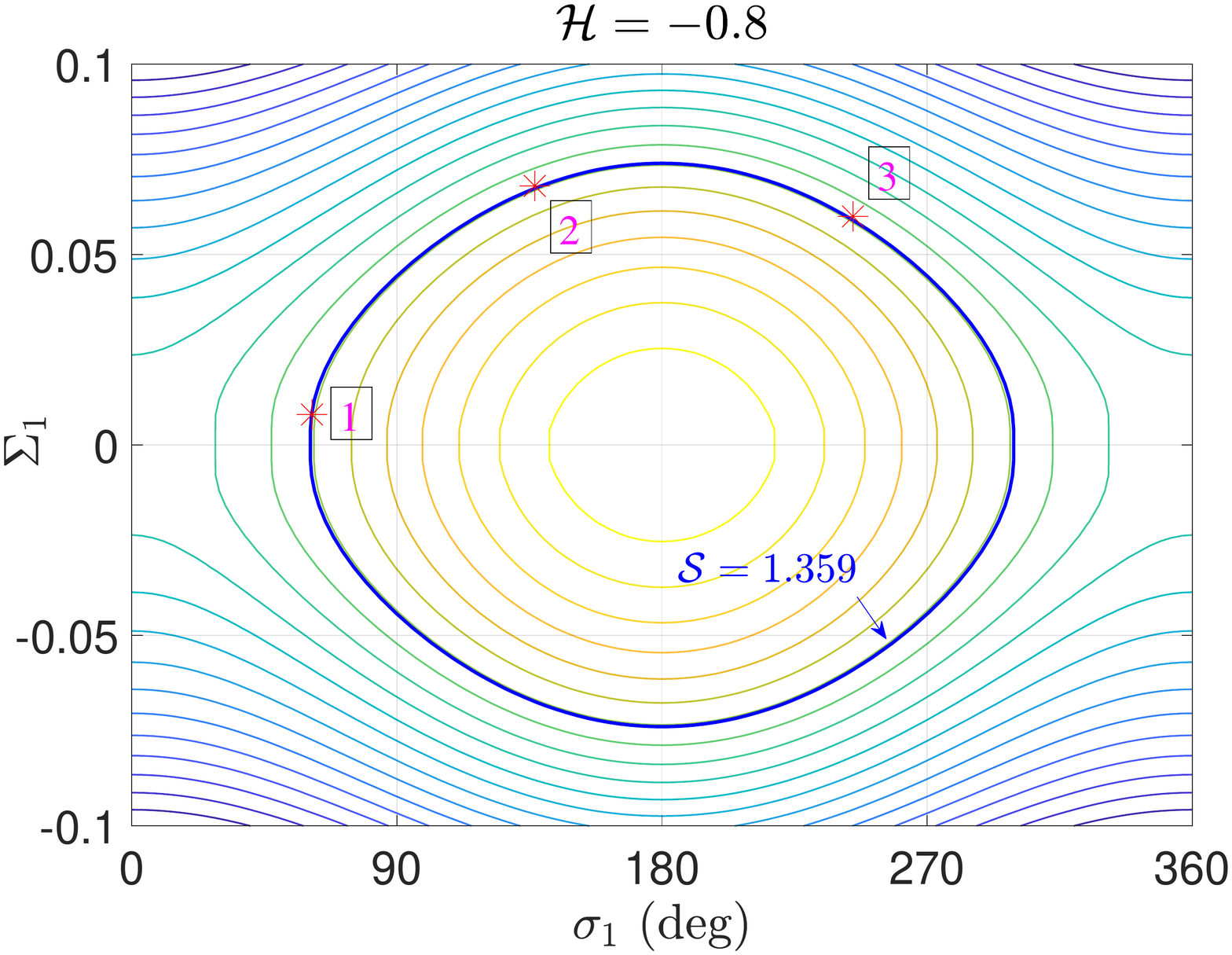}
\includegraphics[width=0.45\textwidth]{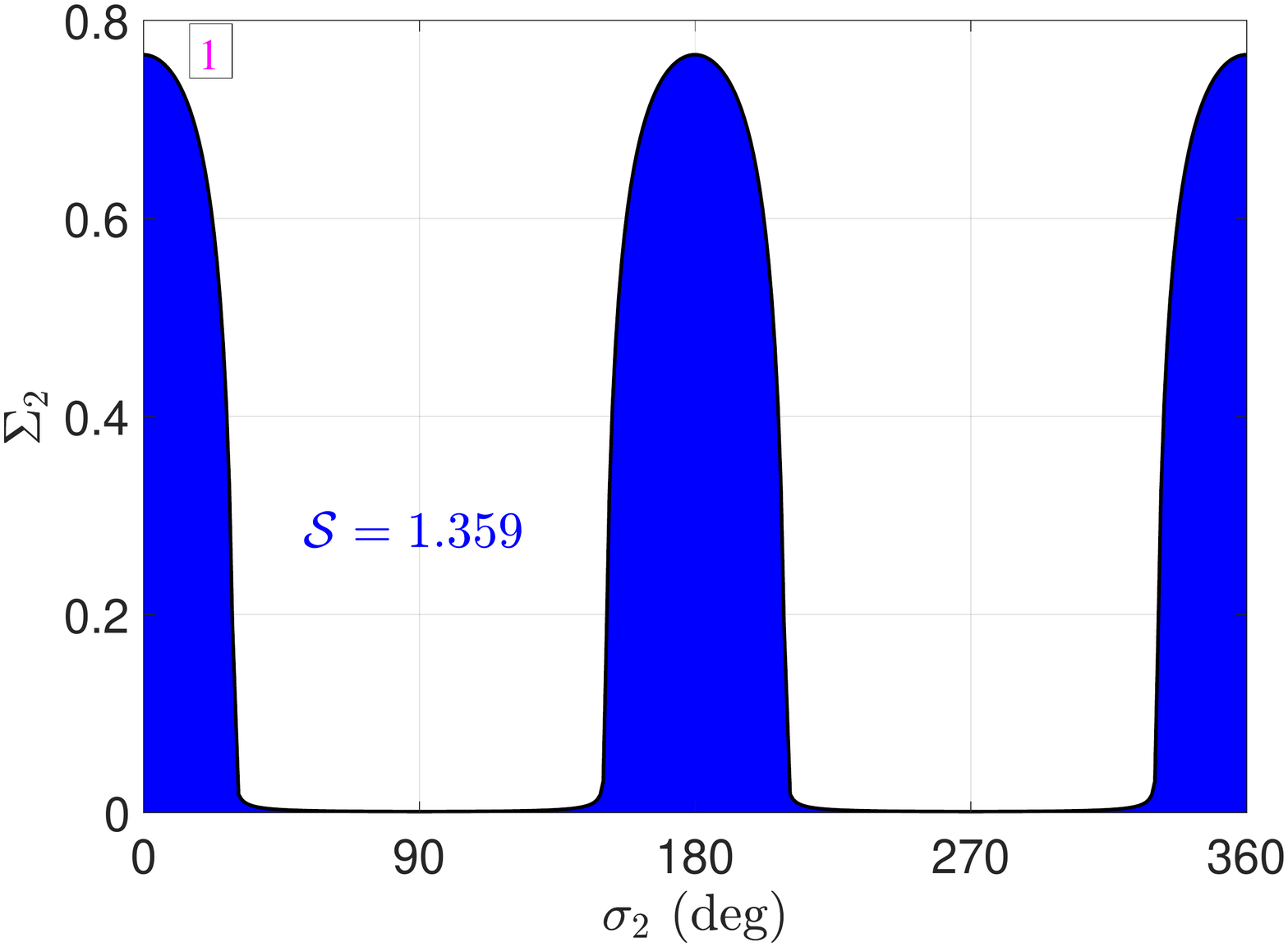}\\
\includegraphics[width=0.45\textwidth]{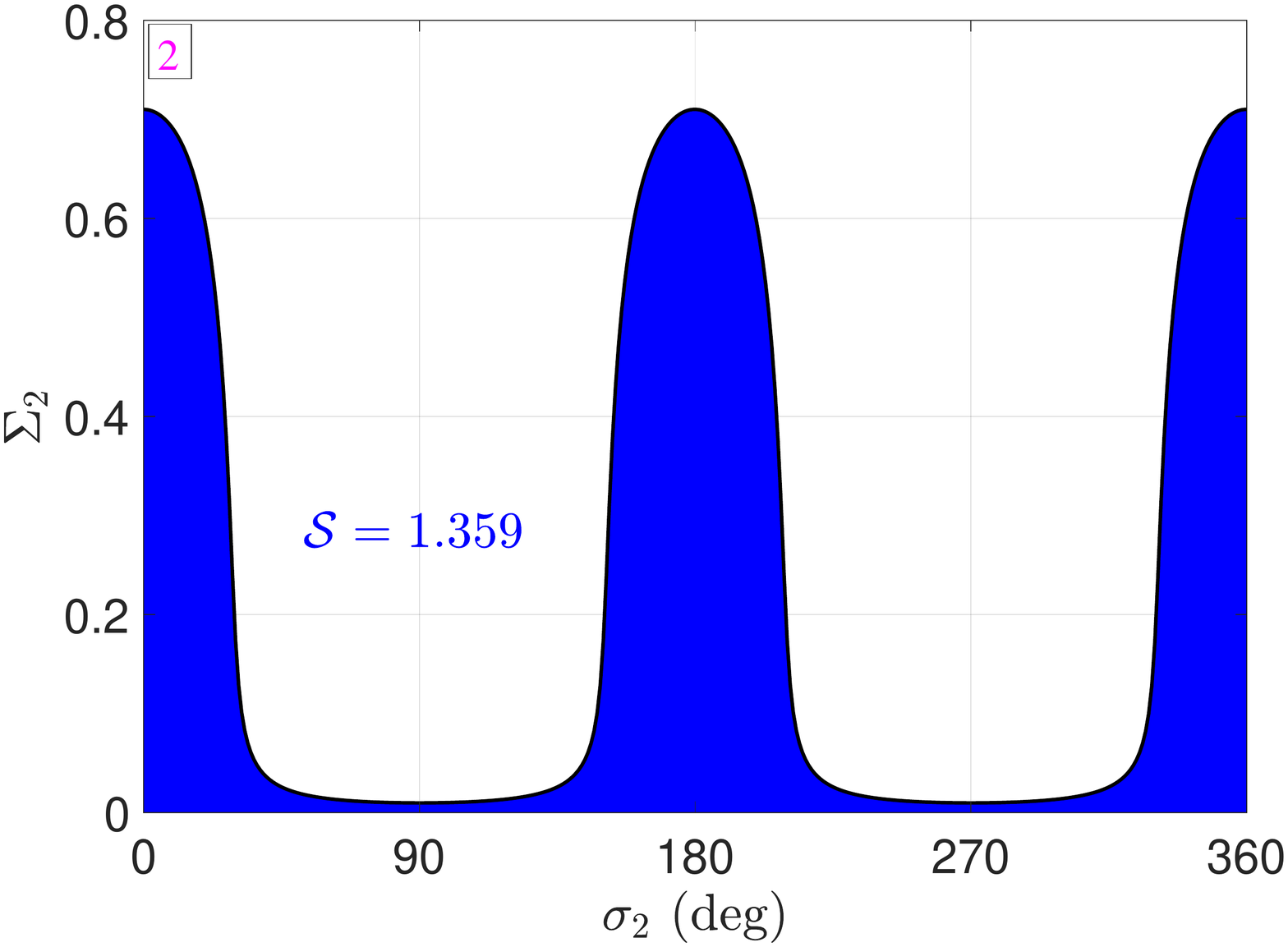}
\includegraphics[width=0.45\textwidth]{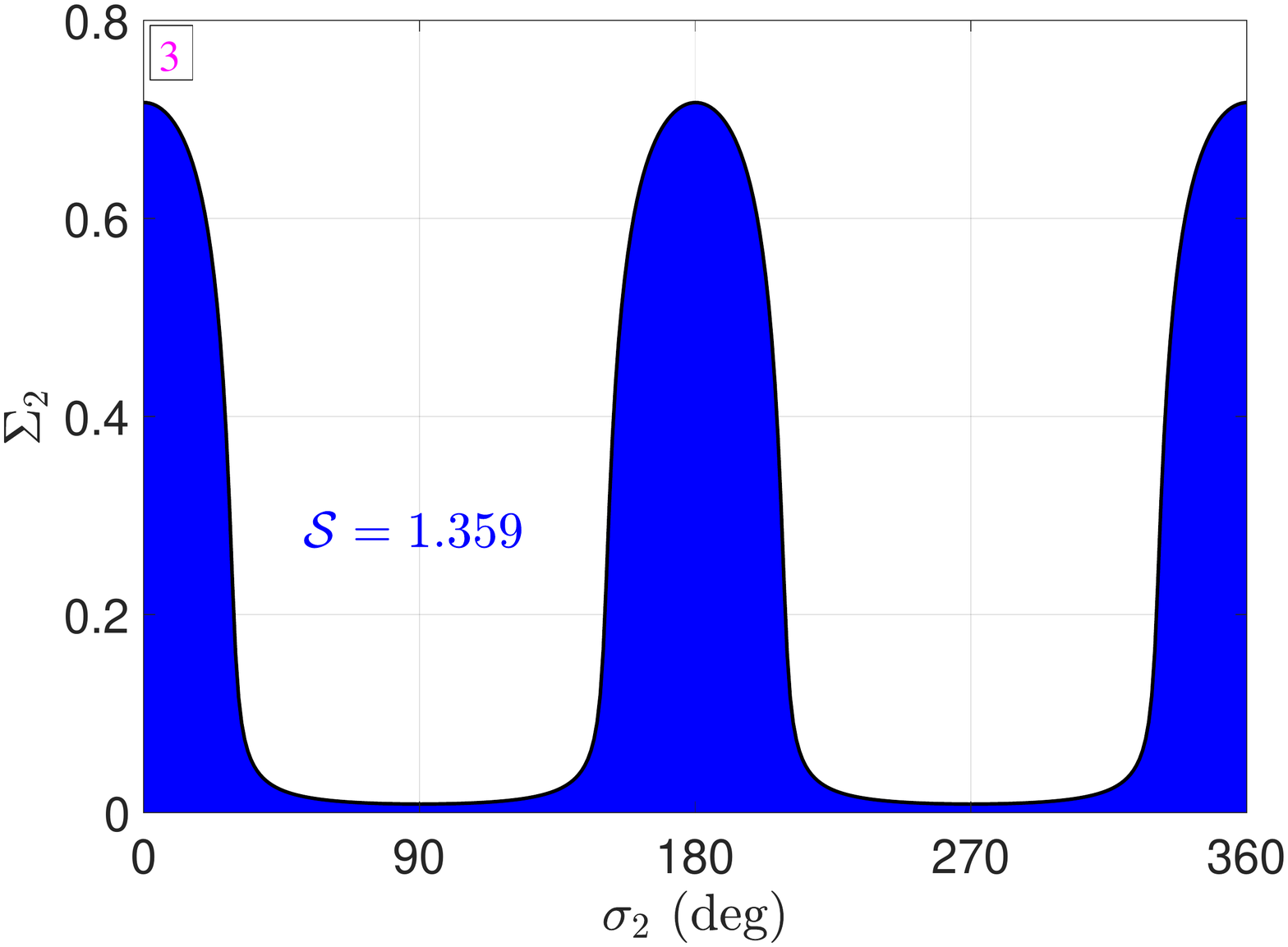}
\caption{Level curves of the adiabatic invariant ${\cal S}$ with given Hamiltonian (\emph{upper-left panel}) and three examples located at the isoline with ${\cal S}=1.359$ (\emph{the rest three panel}s). The adiabatic invariant ${\cal S}$ is defined as the area bounded by the solution curve associated with the `fast' degree of freedom $\Sigma_2(\sigma_2)$.}
\label{Fig10}
\end{figure*}

In the long-term evolution, there are two quasi-conserved quantities: (i) the Hamiltonian denoted by ${\cal H}$ and (ii) the adiabatic invariant ${\cal S}$. As a result, the original two-degree-of-freedom model becomes integral. The motion of test particles occurs on the level curves of ${\cal H}$ and ${\cal S}$ embedded in the four-dimensional phase space $(\sigma_1,\Sigma_1,\sigma_2,\Sigma_2)$. Furthermore, when the Hamiltonian ${\cal H}$ is given, the motion of test particles takes place on the level curves of ${\cal S} ({\cal H}, \sigma_1, \Sigma_1)$. Naturally, the phase portraits can be produced by plotting level curves of ${\cal S}$ in the $(\sigma_1, \Sigma_1)$ space with given Hamiltonian ${\cal H}$ \citep{wisdom1985perturbative, henrard1989motion, henrard1990semi, neishtadt2004wisdom, sidorenko2014quasi, sidorenko2018dynamics, sidorenko2018eccentric, efimov2020analytically}. The resulting phase portraits can be used to reveal dynamical structures in the phase space.

Please refer to the first panel of Fig. \ref{Fig10} for the phase portrait with ${\cal H} = -0.8$. It is observed that there is an island of libration, which is centered at $(\sigma_1 = \pi, \Sigma_1 = 0)$. In particular, a level curve with ${\cal S} = 1.359$ is shown by a blue line and three points on it are marked and denoted by numbers 1, 2 and 3. The associated definitions of the adiabatic invariant ${\cal S}$ for points `1', `2', and `3' are provided in the last three panels of Fig. \ref{Fig10}. Along a certain level curve of ${\cal S}$, the solution curves in the $(\sigma_2,\Sigma_2)$ space are changed.

\begin{figure*}
\centering
\includegraphics[width=0.45\textwidth]{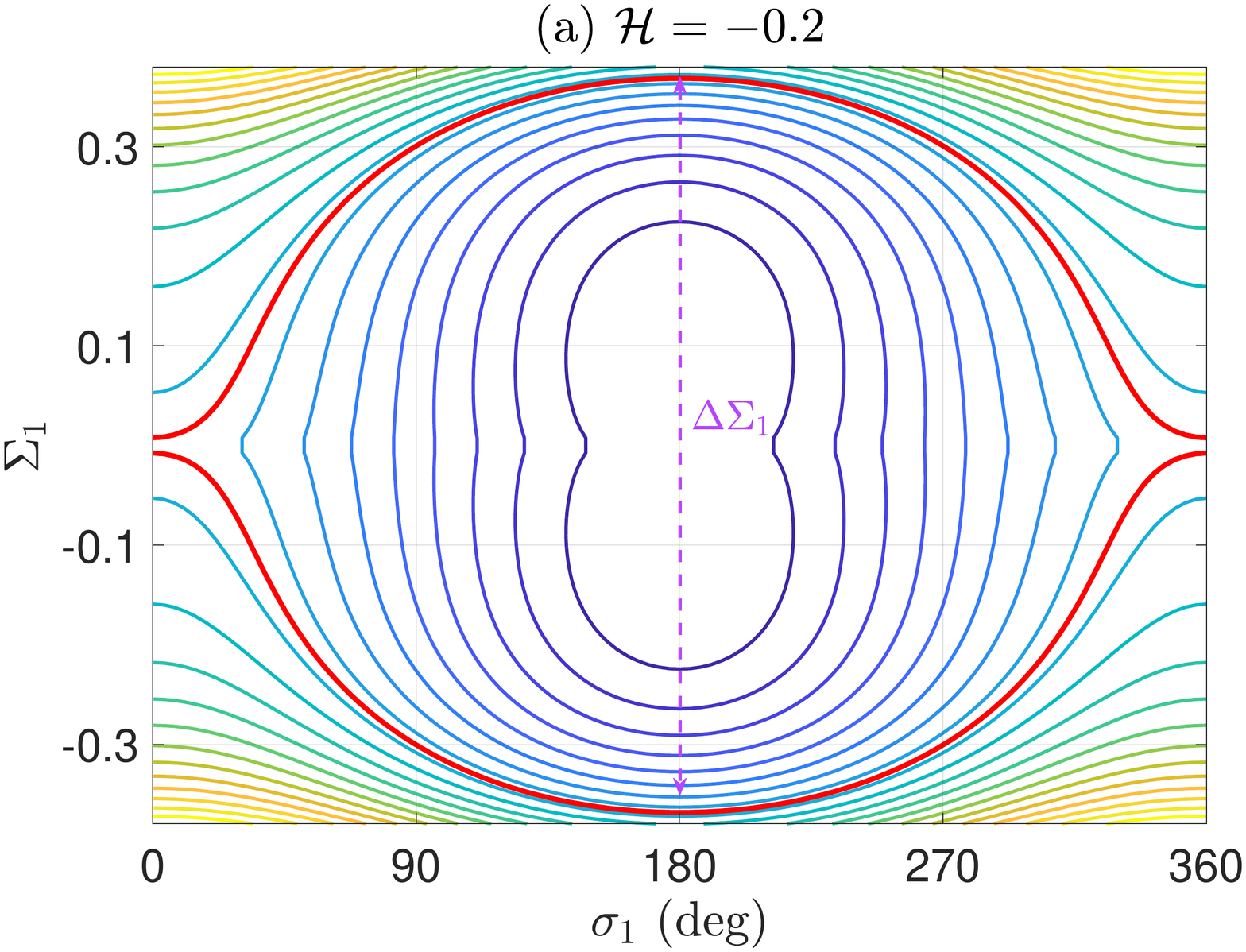}
\includegraphics[width=0.45\textwidth]{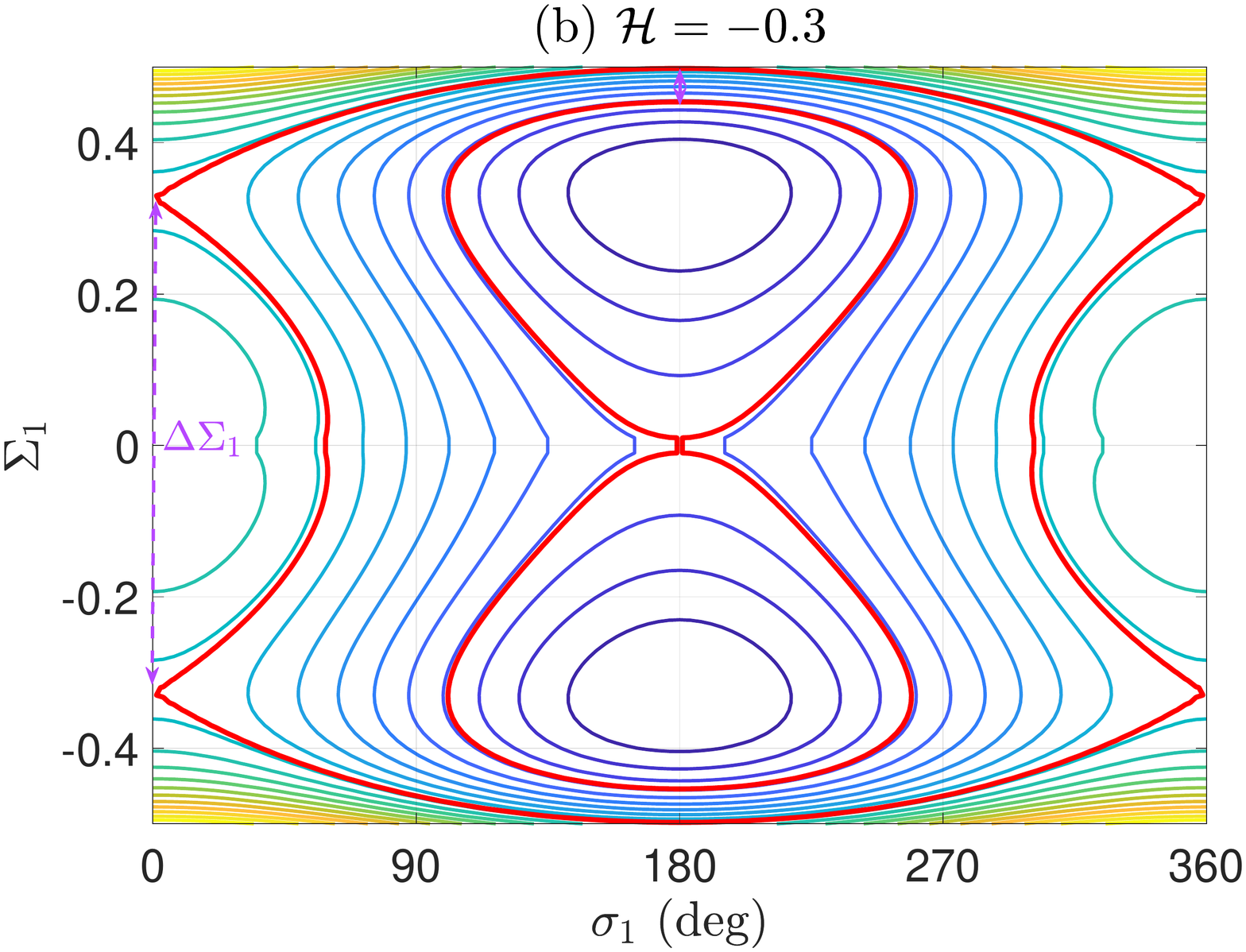}\\
\includegraphics[width=0.45\textwidth]{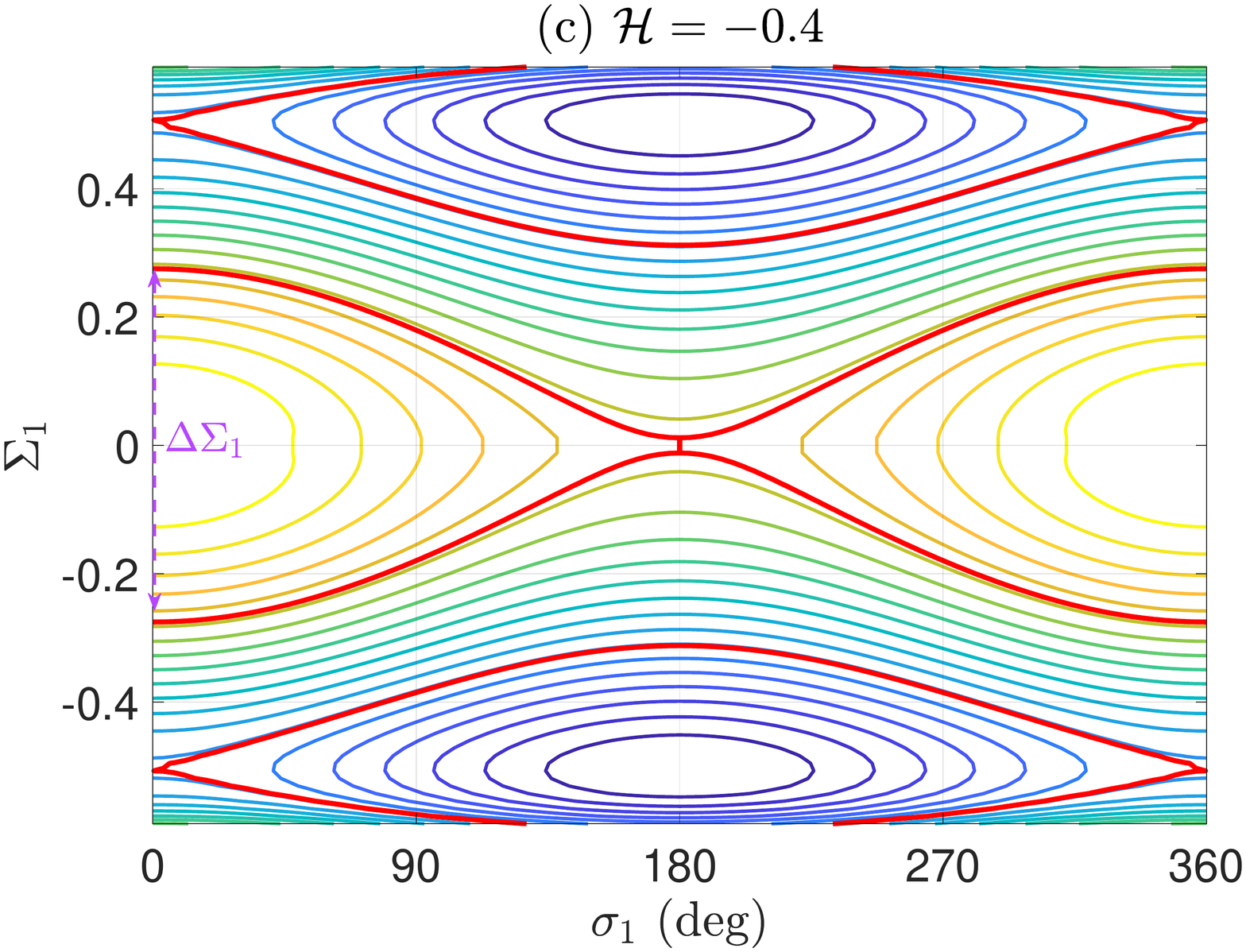}
\includegraphics[width=0.45\textwidth]{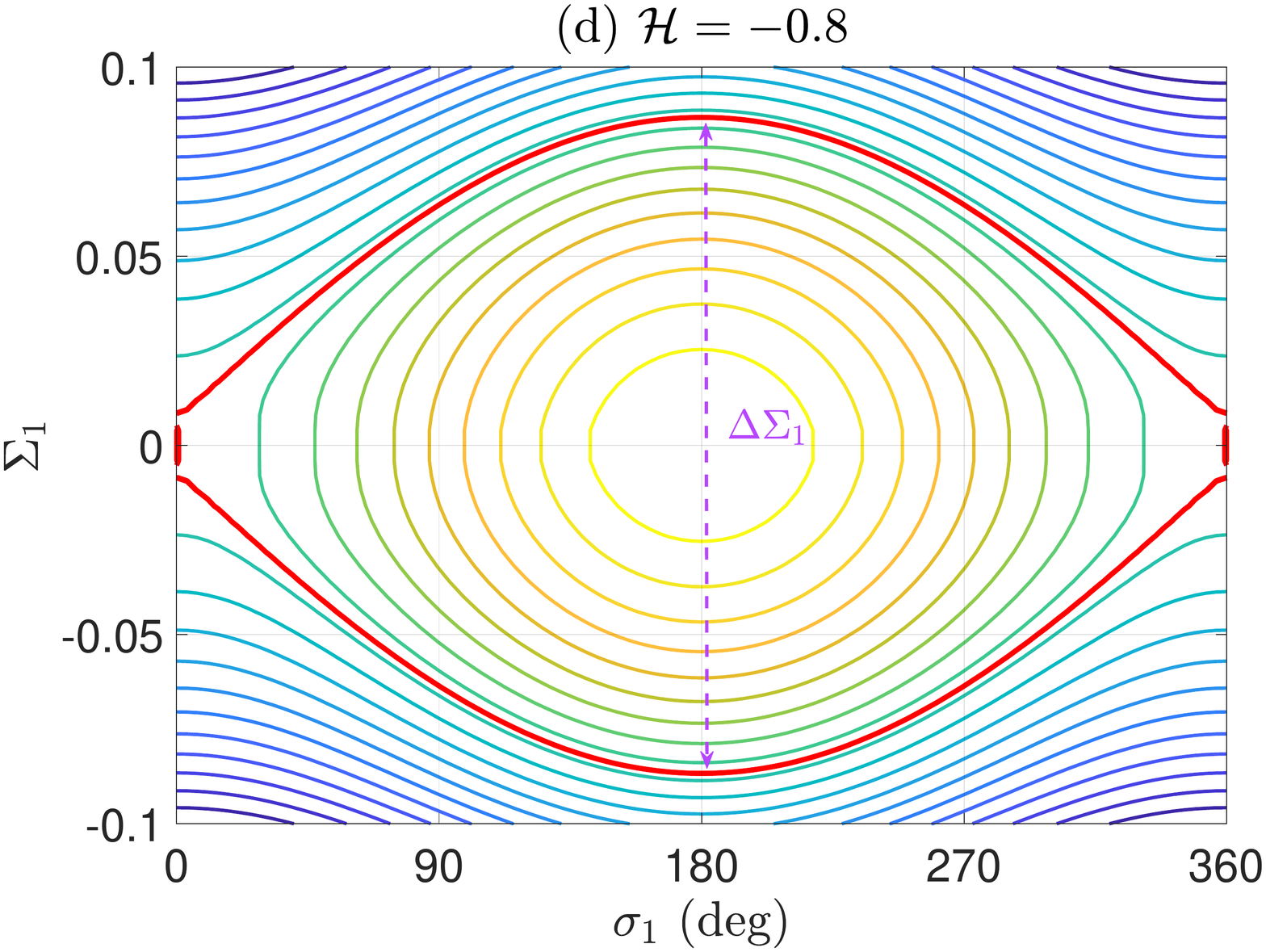}
\caption{Phase portraits (level curves of adiabatic invariant) at different levels of Hamiltonian (${\cal H} = -0.2, -0.3, -0.4, -0.8$). $\Delta \Sigma_1$ measures the size of flipping region. The red lines stand for the dynamical separatrices, dividing circulation regions from libration ones.}
\label{Fig11}
\end{figure*}

Figure \ref{Fig11} presents phase portraits (level curves of ${\cal S}$) for different levels of Hamiltonian (${\cal H} = -0.2, -0.3, -0.4, -0.8$). For convenience of comparison, the magnitudes of Hamiltonian taken here are in accordance with the ones of Poincar\'e surfaces of section shown in Fig. \ref{Fig3}. It is observed that all these four phase portraits are symmetric with respect to the line of ${\Sigma_1 = 0}$. The level curves passing through saddle points are marked by red lines, which play the role of dividing the entire phase space into regions of circulation and libration. Due to the definition of $\Sigma_1 = H = G\cos{i}$, the line of $\Sigma_1 = 0$ stand for the location of polar orbit (i.e., $i = 90^{\circ}$). Thus, the trajectories in those islands of libration centered at $\Sigma_1 = 0$ can flip from prograde to retrograde and back again. In the following discussions, we mainly concentrate on those islands causing orbit flips.

When the Hamiltonian is at ${\cal H} = -0.2$ (see the up-left panel of Fig. \ref{Fig11}), there is a single island of libration in the phase portrait and the resonant center is at $(\sigma_1 = \pi, \Sigma_1 = 0)$. The orbits inside the island of libration can flip.

When the Hamiltonian is decreased to ${\cal H} = -0.3$ (see the up-right panel of Fig. \ref{Fig11}), the dynamical structures in the phase portrait becomes complicated due to the bifurcation of equilibrium points. There are three islands of libration: one is centered at $(\sigma_1 = 0, \Sigma_1 = 0)$ and the other two are symmetric with respect to the line of $\Sigma_1 = 0$. There are two pairs of dynamical separatrices: inner and outer ones shown by red lines. The orbits inside the island centered at $(\sigma_1 = 0, \Sigma_1 = 0)$ and the orbits between the inner and outer separatrices can flip.

When the Hamiltonian is further decreased to ${\cal H} = -0.4$ (see the bottom-left panel of Fig. \ref{Fig11}), there are also three islands of libration: one is centered at $(\sigma_1 = 0, \Sigma_1 = 0)$ and the other two are symmetric with respect to the line of $\Sigma_1 = 0$. The orbits inside the island of libration centered at $(\sigma_1 = 0, \Sigma_1 = 0)$ can flip.

When the Hamiltonian is ${\cal H} = -0.8$, the structures in the phase portrait become simple (see the bottom-right panel of Fig. \ref{Fig11}): there is a single island of libration which is centered at $(\sigma_1 = \pi, \Sigma_1 = 0)$. All the orbits inside the island of libration can flip.

Comparing Fig. \ref{Fig11} with Fig. \ref{Fig3}, we can see that the dynamical structures (islands of libration, stable and unstable equilibrium points, dynamical separatrices) in the phase portraits and Poincar\'e surfaces of section are similar at a certain level of Hamiltonian. Comparisons between Poincar\'e surfaces of section and phase portraits can be found in \citet{henrard1989motion} and \citet{yokoyama1996simple} for the problem of mean motion resonances in the planar elliptic restricted three-body system.

About the problem of orbit flips, \citet{sidorenko2018eccentric} produced similar phase portraits (see Fig. 9 in his work) by taking a different approach based on the adiabatic invariant approximation \citep{henrard1986perturbation, henrard1990semi, henrard1993adiabatic}.

\begin{figure*}
\centering
\includegraphics[width=0.45\textwidth]{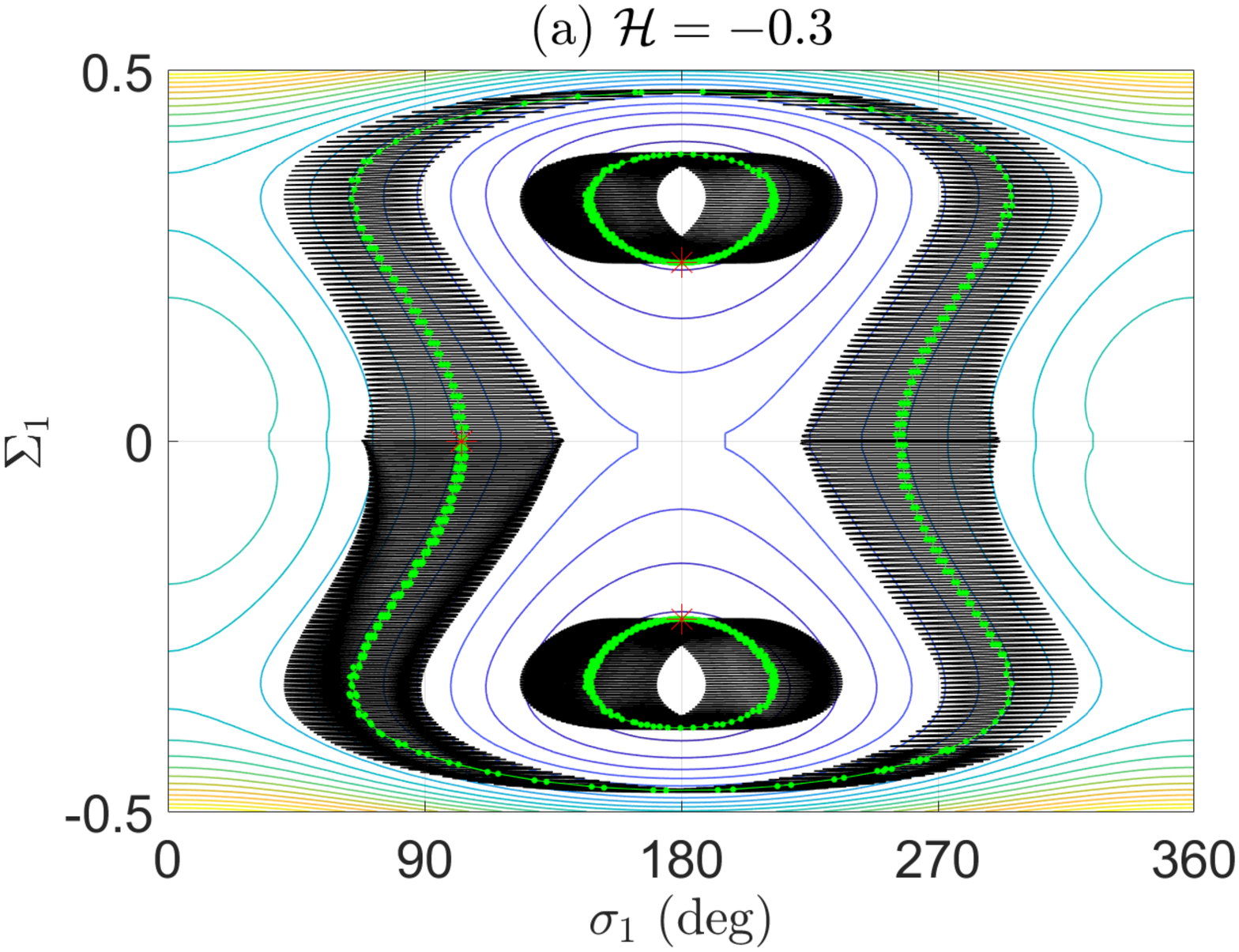}
\includegraphics[width=0.45\textwidth]{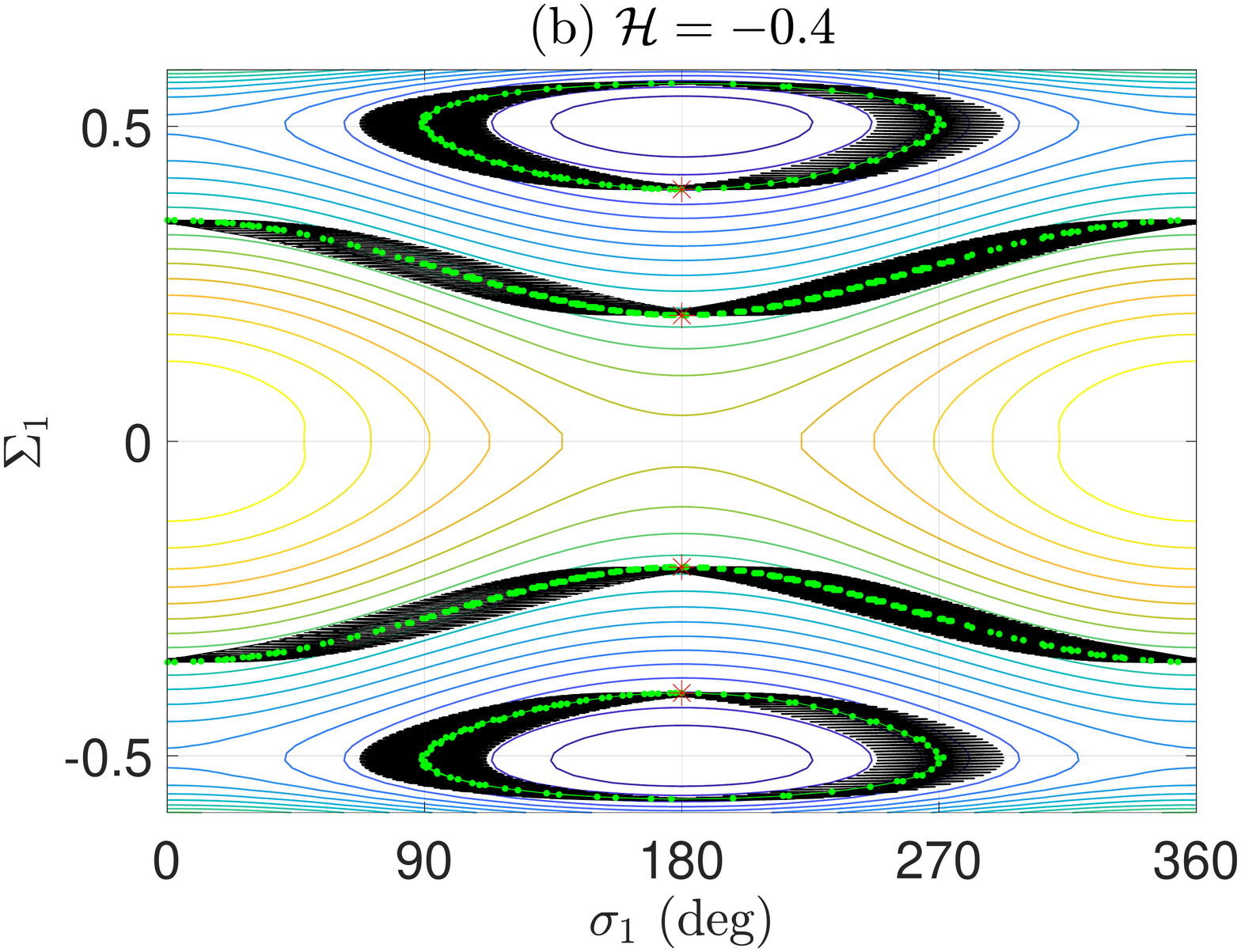}
\caption{Phase portraits (similar to the ones shown in Fig. \ref{Fig11}) as well as numerically propagated trajectories shown by black lines. The starting points of numerical trajectories are marked by red stars and the initial angle $\sigma_2(0)$ is set as zero for all numerical trajectories. The initial state $\Sigma_2 (0)$ can be produced from the given Hamiltonian. The green dots stand for the points on the numerical trajectories holding $\sigma_2 (=\omega) = 0$. It is observed that the green points on the section follow closely along the level curves shown in the phase portraits.}
\label{Fig12}
\end{figure*}

To validate the phase portraits, we numerically integrate the equations of motion for the original two-degree-of-freedom problem and show the resulting trajectories in the $(\sigma_1,\Sigma_1)$ space. Figure \ref{Fig12} reports phase portraits for the Hamiltonian at ${\cal H} = -0.3, -0.4$ (the same as the up-right and bottom-left panels of Fig. \ref{Fig11}) as well as several numerically propagated trajectories shown by black lines. The points on the numerical trajectories with $\omega = 0$ are marked by green dots. We could understand that the green dots represent the points on the Poincar\'e sections defined by $\omega = 0$ for numerical trajectories. It is observed from Fig. \ref{Fig12} that (a) the numerical trajectories propagated under the original two-degree-of-freedom model have periodic oscillations\footnote{The periodic oscillations are removed in perturbation treatments.}, (b) the long-term behaviors of numerical trajectories agree well with the level curves arising in the phase portrait, and (c) the points on the section defined by $\omega = 0$ have an excellent agreement with the level curves of adiabatic invariant. Based on these discussions, we can conclude that the resonant model formulated by means of perturbation treatments and the resulting phase portraits are valid and applicable in predicting the long-term dynamics in the $(\sigma_1,\Sigma_1)$ space.

\begin{figure*}
\centering
\includegraphics[width=0.6\textwidth]{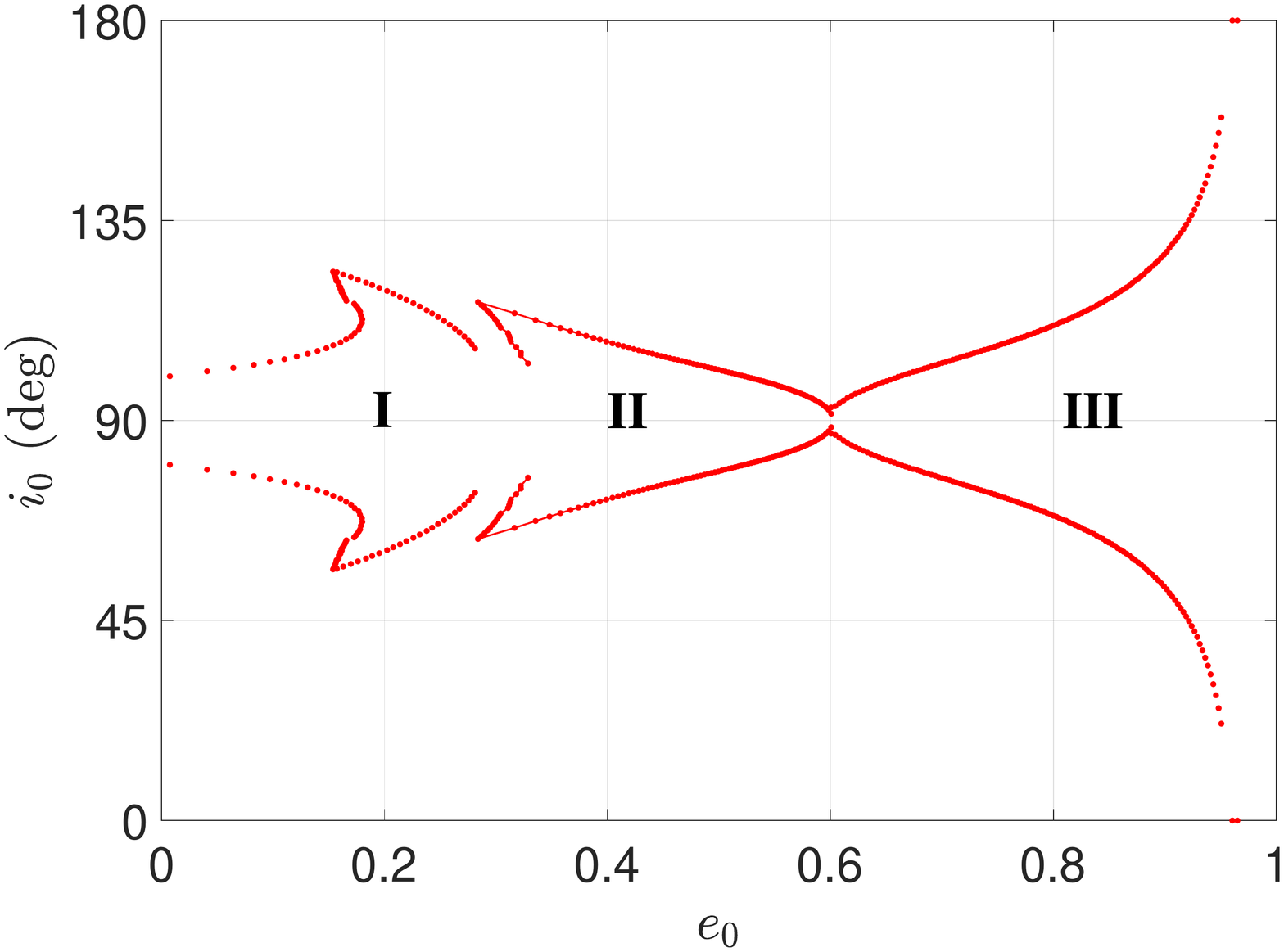}
\caption{Analytical boundaries of flipping regions shown in the $(e_0, i_0)$ space. Similarly, the regions of orbit flips are denoted by I, II and III from the left to right. In particular, the libration center of the critical argument $\sigma_1$ is at $\sigma_1 = \pi$ in regions I and III and it is at $\sigma_1 = 0$ in region II.}
\label{Fig13}
\end{figure*}

From the phase portraits, we can evidently observe that the center of island of libration causing orbit flips is at either zero or $\pi$, depending on the value of Hamiltonian. Thus, according to resonant centers, we can divide the flipping orbits into two types: the first type with the critical argument $\sigma_1$ librating around $\pi$ (type I) and the other one with critical argument $\sigma_1$ librating around zero (type II). It is not difficult to see that these two types correspond to cases I and II discussed in Section \ref{Sect3}. In order to evaluate the size of flipping region, we use $\Delta \Sigma_1 (={\Delta H})$ to measure the resonant width for the island of libration causing orbit flips.

By analysing the size of the island of libration causing orbit flips, it is possible to produce such a function $\Delta \Sigma ({\cal H})$, which stands for the resonant width as a function of the Hamiltonian ${\cal H}$. Due to the symmetry property of phase portraits about the polar line, it is possible to obtain the lower and upper boundaries for orbit flips in the eccentricity--inclination space. The analytical results are reported in Fig. \ref{Fig13}, where the regions are denoted by I, II and III from the left to right. It is noted that in regions I and III the critical argument $\sigma_1$ of flipping orbits is librating around $\pi$ (the resonant center is at $\sigma_1 = \pi$) and, in region II, the critical argument $\sigma_1$ of flipping orbits is librating around zero (the resonant center is at $\sigma_1 = 0$).

\section{Conclusions and discussions}
\label{Sect7}

In this work, the problem of orbit flips caused by the eccentric von Zeipel--Lidov--Kozai effect is systematically investigated under the test particle limit at the octupole-order approximation. By integrating the secular equations of motion, the flipping regions in the whole $(e_0, i_0)$ space are produced by taking the initial angles at $\omega_0 = 0$ and $\Omega_0=\pi$ (case I) and $\omega_0 = 0$ and $\Omega_0=0$ (case II). The regions corresponding to case I are distributed in the low-eccentricity and high-eccentricity spaces and the regions corresponding to case II are distributed in the intermediate-eccentricity space. The results cover both the low-eccentricity high-inclination (LeHi) case and the high-eccentricity low-inclination (HeLi) case, which are discussed in \citet{li2014eccentricity}. Numerical simulations show that (a) there are three distinct regions, denoted by I (low-eccentricity region), II (intermediate-eccentricity region) and III (high-eccentricity region) from the left to right, and (b) the structures of flipping areas remain qualitatively similar under dynamical models with different magnitudes of $\epsilon$. In particular, when the eccentricity is higher than a critical value $e_c (\epsilon)$, orbit flips can take place from nearly-coplanar configurations (corresponding to the HeLi case where flips can occur with nearly coplanar configurations).

In order to understand the numerical structures of flipping regions arising in the $(e_0,i_0)$ space, the problem of orbit flips caused by the eccentric von Zeipel--Lidov--Kozai effect is investigated by means of three approaches: Poincar\'e sections and dynamical system theory (periodic orbits and invariant manifolds) and perturbation treatments. It should be mentioned that the solution curve $\Sigma_2 (\sigma_2)$ and the adiabatic invariant ${\cal S}$ are numerically evaluated during the process of perturbative treatments. From the viewpoint of Poincar\'e sections, orbit flips are due to the existence of the islands of libration centered at $H = 0$. From the viewpoint of dynamical system theory (periodic orbits and invariant manifolds), orbit flips are due to the existence of polar periodic orbits and invariant manifolds. From the viewpoint of resonance, orbit flips are due to the libration of the critical argument $\sigma_1 = h +{\rm sign}(H)g$, which supports the interpretation of the eccentric von Zeipel--Lidov--Kozai effect made by \citet{sidorenko2018eccentric}. Based on these discussions, we can summarize the essence of flipping orbits as follows:
\begin{itemize}
  \item Flipping orbits are a kind of quasi-periodic trajectories around the stable polar periodic orbits. Invariant manifolds emanating from the unstable polar periodic orbits at the same level of Hamiltonian provide the boundaries of quasi-periodic orbits in phase spaces.
  \item Flipping orbits are a kind of resonant trajectories. Resonant trajectories with resonant center at $\sigma_1 = \pi$ corresponds to type I (or case I) and the ones with resonant center at $\sigma_1 = 0$ correspond to type II (or case II). Resonant width stands for the size of flipping region.
\end{itemize}

\begin{figure*}
\centering
\includegraphics[width=0.6\textwidth]{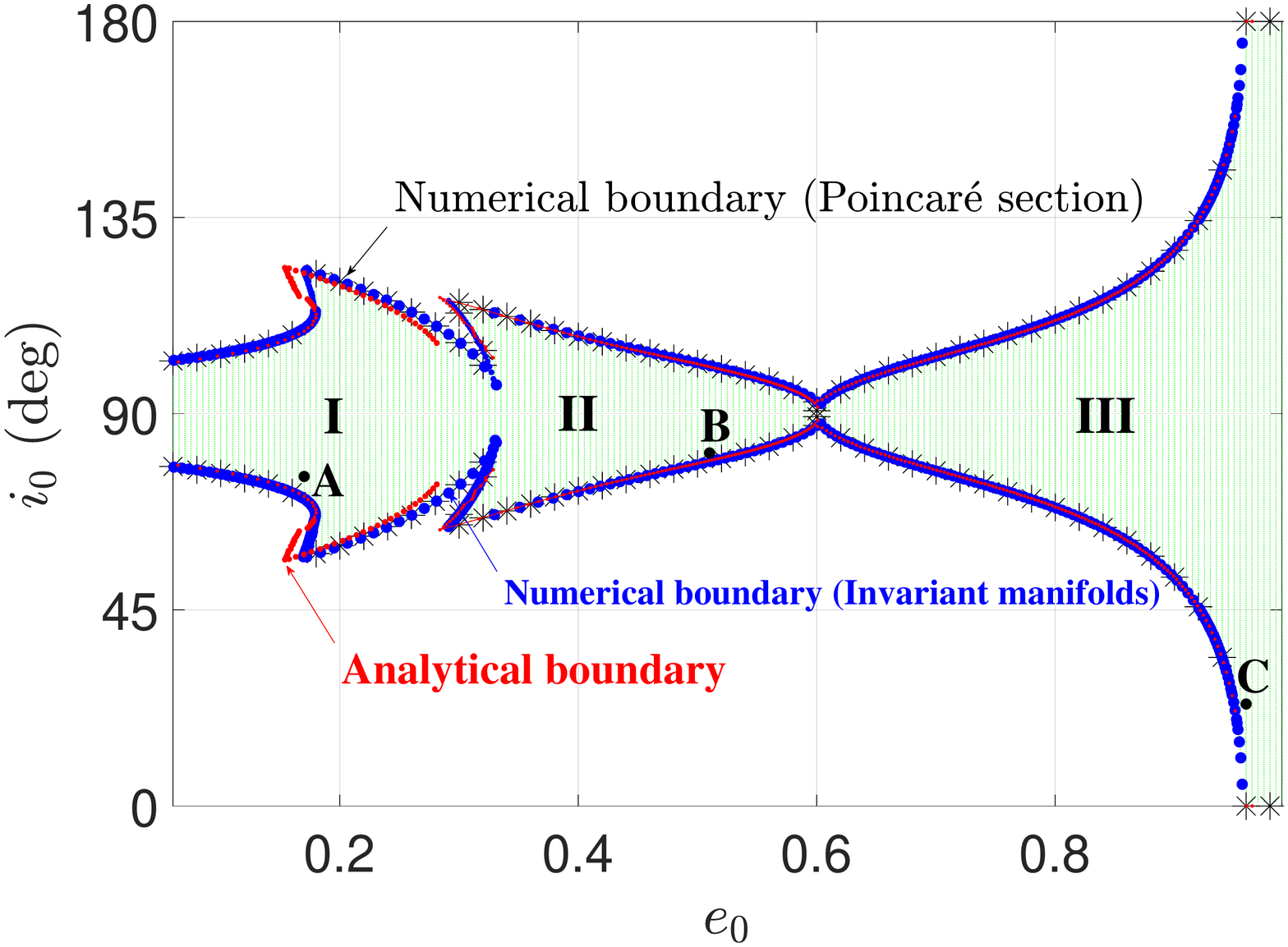}
\caption{Comparisons of flipping boundaries produced by three approaches. The black stars stand for the numerical boundary based on Poincar\'e surfaces of section, the blue dots stand for the numerical boundary based on invariant manifolds of unstable polar orbits and the red dots stand for the analytical boundary obtained by means of perturbative treatment. Three examples are marked by `A' (in region I), `B' (in region II) and `C' (in region III) and their evolution of elements are given in Fig. \ref{Fig15}. The shaded regions denoted by `I', `II' and `III' are the same as the left one of Fig. \ref{Fig1}.}
\label{Fig14}
\end{figure*}

In Fig. \ref{Fig14}, we report all the numerical and analytical results together. The black stars stand for the numerical boundaries of orbit flips obtained by analysing Poincar\'e sections, the green dots represent the numerical boundaries of orbit flips produced by analysing invariant manifolds, and the red dots correspond to the analytical boundaries of orbit flips generated by perturbation treatments. In addition, the green shaded regions stand for the distribution of flipping orbits in the $(e_0,i_0)$ space, which is the same as the left panel of Fig. \ref{Fig1}. Similarly, the regions of orbit flips are denoted by I, II and III from the left to right. From Fig. \ref{Fig14}, it is observed that (a) the boundaries of orbit flips are symmetric with respect to $i_0 = 90^{\circ}$ due to the symmetry of the Hamiltonian, (b) the numerical boundaries of orbit flips present excellent agreement with each other, and (c) the analytical boundary agrees well with the numerical boundaries.

\begin{figure*}
\centering
\includegraphics[width=0.98\textwidth]{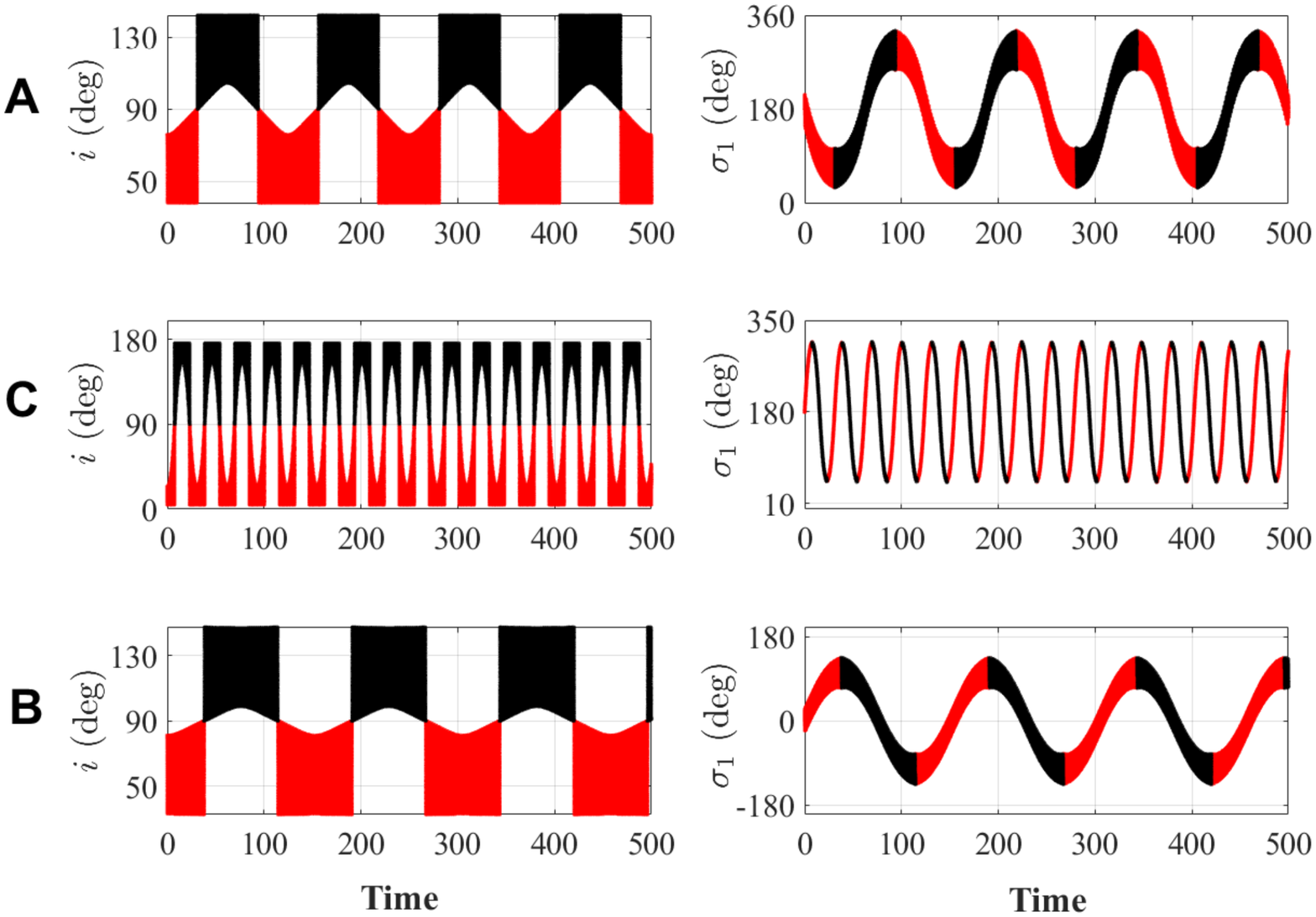}
\caption{Time histories of inclination $i$ and resonant argument $\sigma_1$ for three representatives of flipping orbits shown in Fig. \ref{Fig14}. For convenience, the points with $i<90^{\circ}$ on the numerical trajectories are shown in red and the points with $i>90^{\circ}$ are shown in black. For trajectories `A' (in region I) and `C' (in region III), the critical argument $\sigma_1$ librates around $180^{\circ}$ and, for trajectory `B' (in region II), the critical argument $\sigma_1$ librates around $0^{\circ}$.}
\label{Fig15}
\end{figure*}

Three examples are marked in Fig. \ref{Fig14} and they are denoted by letters `A', `B' and `C'. Orbit `A' is located in region I (low-eccentricity region), orbit `B' is located in region II (intermediate-eccentricity region) and orbit `C' is located in regions III (high-eccentricity region). The orbits are numerically propagated over 500 dimensionless units of time and their temporal evolutions of inclination $i$ and resonant argument $\sigma_1$ are reported in Fig. \ref{Fig15}. For convenience, the prograde arcs are shown in red and the retrograde arcs are shown in black. As expected, the resonant arguments of orbit `A' and `C' are librating around $\sigma_1 = \pi$ (flipping orbits `A' and `C' belong to type I or case I), and the resonant argument of orbit `B' is librating around $\sigma_1 = 0$ (flipping orbit `B' belongs to type II or case II).

\begin{acknowledgments}
Hanlun Lei thanks Dr. Jian Li for helpful discussions in preparing the manuscript and an anonymous reviewer for suggestions that improve the quality of this manuscript. This work is supported by the National Natural Science Foundation of China (Nos 12073011).
\end{acknowledgments}

\appendix
\section{Discussion about $\Omega_e$ and $\sigma_1$}
\label{A1}

\begin{figure}
\centering
\includegraphics[width=0.5\textwidth]{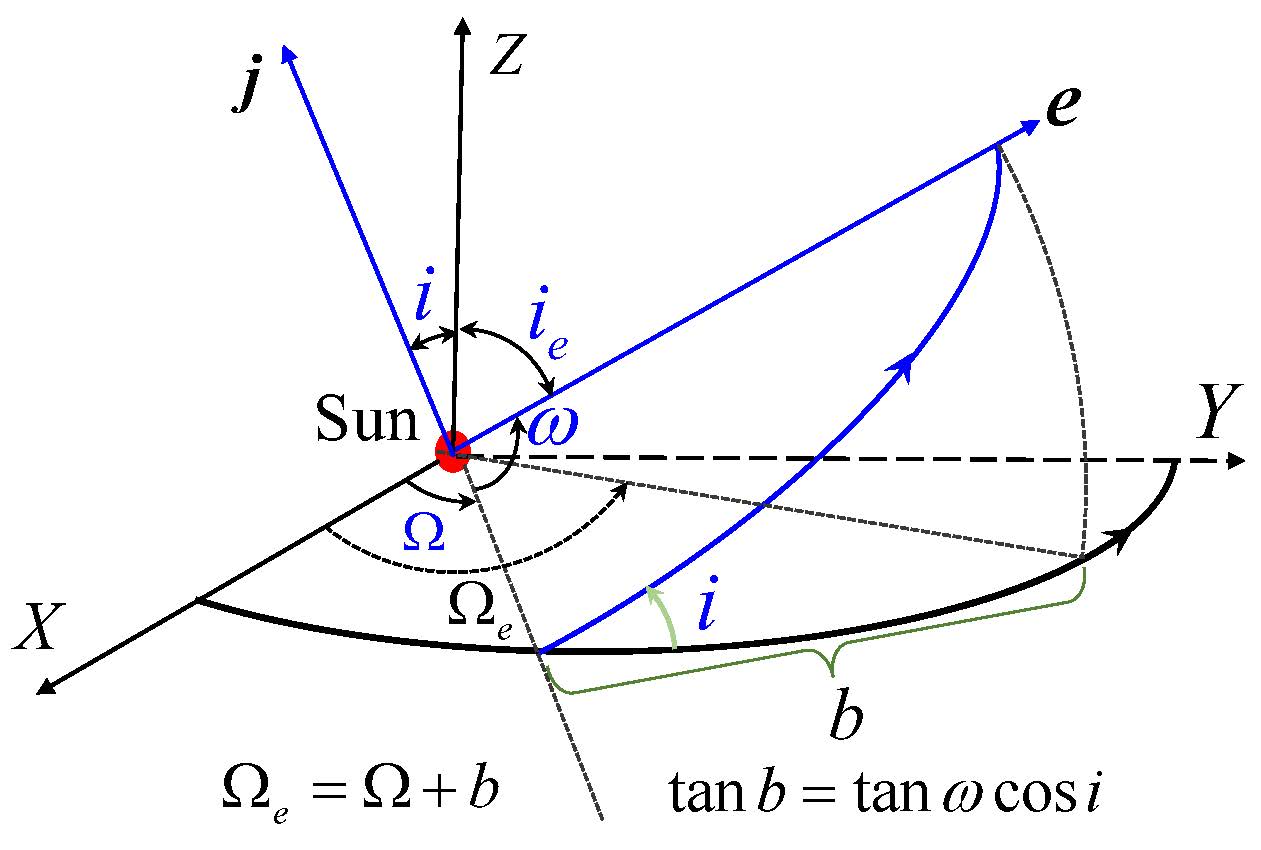}
\caption{Schematic diagram for the definition of $i_e$ and $\Omega_e$ adopted by \citet{katz2011long}.}
\label{Fig_A1}
\end{figure}

The colatitudinal angle $i_e$ and longitude $\Omega_e$ are introduced by \citet{katz2011long} to express the eccentricity vector of test particle's orbit by
\begin{equation*}
{\bm e} = e \left(\sin{i_e}\cos{\Omega_e}, \sin{i_e}\sin{\Omega_e}, \cos{i_e}\right).
\end{equation*}
Please refer to Fig. \ref{Fig_A1} for the geometrical definition. We can get that the angles $i_e$ and $\Omega_e$ are given by
\begin{equation*}
\cos{i_e} = \sin{\omega}\sin{i},\quad \Omega_e = \Omega + \arctan(\tan{\omega}\cos{i}).
\end{equation*}
According to \citet{katz2011long}, the equation of $\Omega_e$ under the quadrupole order reads
\begin{equation*}
{\dot \Omega}_e = \frac{3}{8} \sqrt{1-e^2}\cos{i}\left(8- \frac{6}{\sin^2{i_e}}\right),
\end{equation*}
which shows that $\Omega_e$ is slowly varying when $(1-e^2)\cos^2{i} \ll 1$. In particular, ${\dot \Omega}_e$ is equal to zero when the inclination is at $i=90^{\circ}$. As stated by \citet{katz2011long}, the longitude $\Omega_e$ is useful to describe the very long-term behavior for the system at hand.

In the current work, the critical argument $\sigma_1 = h + {\rm sign}(H)g = \Omega + {\rm sign}(\cos{i})\omega$ is adopted to describe the long-term behaviors. In particular, it is $\sigma_1 = \Omega + \omega$ for prograde configurations and it is $\sigma_1 = \Omega - \omega$ for retrograde configurations. When the configuration is near to the coplanar case (i.e., $i$ is close to $0$ or $\pi$), we can see that the angle $\Omega_e$ can be approximated by $\Omega_e \approx \Omega + {\rm sign}(\cos{i})\omega$. That is to say, the longitude $\Omega_e$ adopted in \citet{katz2011long} is consistent with the critical argument $\sigma_1$ taken in this work for nearly coplanar configurations.

According to the discussions made by \citet{shevchenko2016lidov} and \citet{lei2021dynamical}, the longitude of pericenter is generally defined by $\varpi = \Omega + {\rm sign}(\cos{i})\omega$. Thus, the critical argument $\sigma_1$ adopted in this work is equal to the longitude of pericenter $\varpi$ for both prograde and retrograde configurations. Considering the fact that the longitude of pericenter for the perturber's orbit is set as $\varpi_p = 0$ due to the choice of the $x$-axis of the coordinate frame, we can further write the critical argument $\sigma_1$ as $\sigma_1 = \varpi = \varpi - \varpi_p $, implying that $\sigma_1$ taken in this work is the critical argument of apsidal resonance in the test-particle limit. As a result, we can conclude that the phenomenon of orbit flips is due to the apsidal resonance under the octupole-level approximation.

Next, let us discuss the critical argument $\sigma_1$ and its time derivative under the quadrupole-level approximation (at the leading order). Up to the second order in semimajor axis ratio, the Hamiltonian is
\begin{equation*}
{{\cal H}_2} = \frac{1}{2}\left( {1 - {G^2}} \right) - \frac{{{H^2}}}{{{G^2}}} - \frac{3}{2}\left( {\frac{1}{{{G^2}}} - 1} \right){H^2} - \frac{5}{2}\left( {1 - {G^2} + {H^2} - \frac{{{H^2}}}{{{G^2}}}} \right)\cos \left( {2g} \right)
\end{equation*}
and the time derivatives of $g$ and $h$ can be written as
\begin{equation*}
\dot g = \frac{{\partial {{\cal H}_2}}}{{\partial G}},\quad \dot h = \frac{{\partial {{\cal H}_2}}}{{\partial H}}.
\end{equation*}
Thus under the quadrupole order the time derivative of $\sigma_1$ can be expressed by
\begin{equation*}
{\dot \sigma}_1 = \frac{{\partial {{\cal H}_2}}}{{\partial H}} + {\rm sign}(H) \frac{{\partial {{\cal H}_2}}}{{\partial G}}.
\end{equation*}
Let us average the frequency ${\dot \sigma}_1$ over one period of a rotating Kozai cycle as follows
\begin{equation*}
\left\langle {{{\dot \sigma }_1}} \right\rangle  = \frac{1}{T}\int\limits_0^{T} {\left[ {\frac{{\partial {{\cal H}_2}}}{{\partial H}} + {\rm sign}\left( H \right)\frac{{\partial {{\cal H}_2}}}{{\partial G}}} \right]{\rm d}t}
\end{equation*}
where $T$ is the period of a rotating Kozai cycle. Without loss of generality, the initial state of a certain rotating Kozai cycle is assumed at $\omega_0 = 0$ and the associated eccentricity and inclination are denoted by $e_0$ and $i_0$. Thus, the initial Delaunay's variables are given by $g(0) = 0$, $G(0) = \sqrt{1-e_0^2}$ and $H(0)=G(0)\cos{i_0}$.

\begin{figure}
\centering
\includegraphics[width=0.6\textwidth]{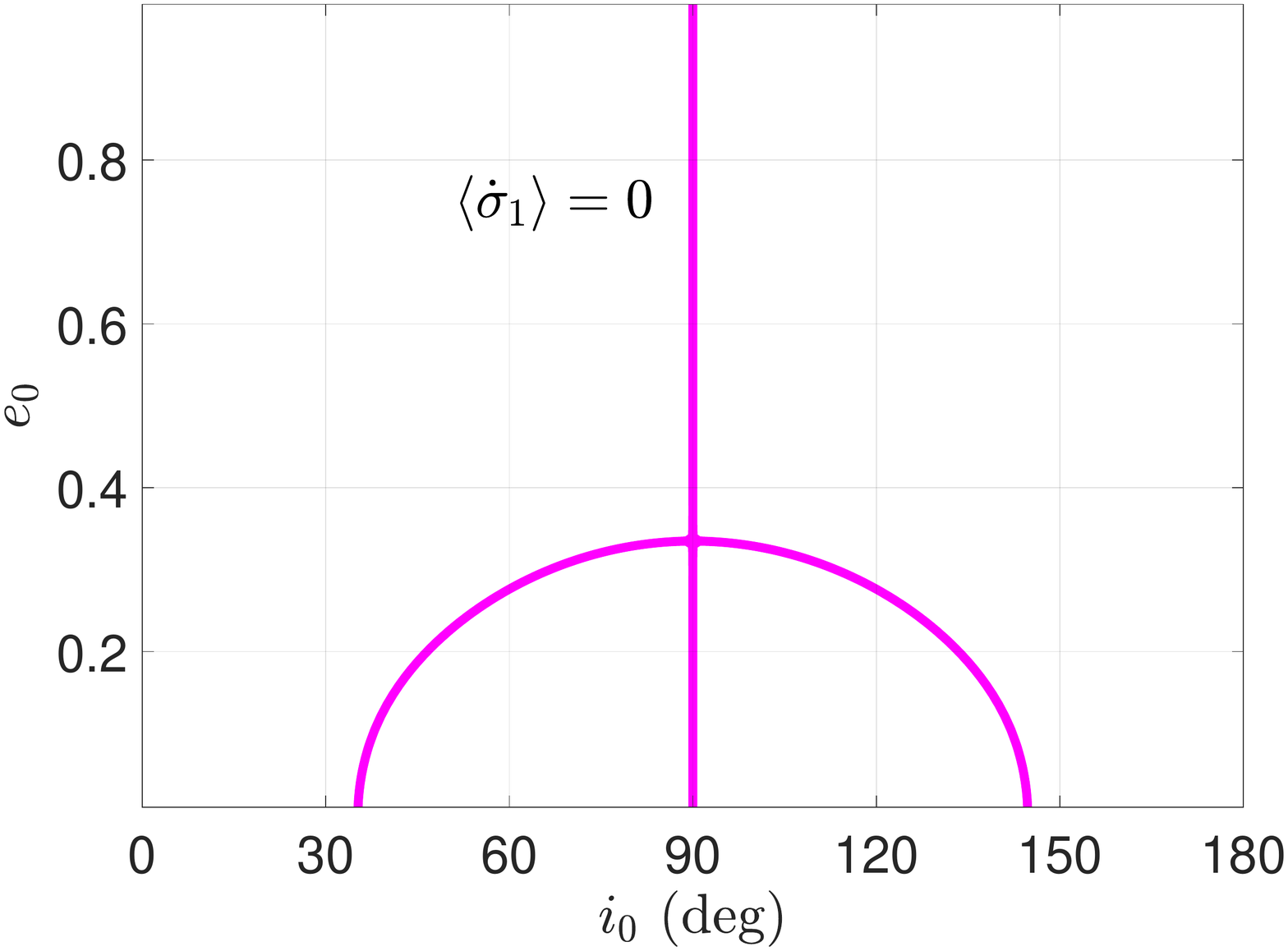}
\caption{The distribution of $\left\langle {{{\dot \sigma }_1}} \right\rangle = 0$ in the initial inclination--eccentricity $(i_0,e_0)$ space. Here $\left\langle {{{\dot \sigma }_1}} \right\rangle$ stands for the average of ${\dot \sigma}_1$ over one period of a rotating Kozai cycle starting from $(\omega_0 = 0,e_0,i_0)$.}
\label{Fig_A2}
\end{figure}

Figure \ref{Fig_A2} shows the solution of $\left\langle {{{\dot \sigma }_1}} \right\rangle = 0$ in the $(i_0,e_0)$ space. It is observed that (a) at the inclination $i_0 = 90^{\circ}$, the average time derivative of $\sigma_1$ (over one period of Kozai cycle) is zero, and (b) in the low-eccentricity space there are another two branches of resonances, corresponding to asymmetric resonances. This distribution of resonance center can be understood with the help of Poincar\'e sections (see Fig. \ref{Fig3}) and phase portraits (see Fig. \ref{Fig11}).

From the aforementioned discussions, we can see that, at the quadrupole order (leading order), it holds ${\dot \Omega}_e = 0$ and $\left\langle {{{\dot \sigma }_1}} \right\rangle = 0$ at the location of $i = 90^{\circ}$. Thus, we can state that $\Omega_e$ and $\sigma_1$ are slow variables for the considered cases and both of them are applicable for describing the very long-term behaviors caused by eccentric von Zeipel--Lidov--Kozai effects.



\bibliography{mybib}{}
\bibliographystyle{aasjournal}



\end{document}